\documentclass[12pt]{JHEP3}

\usepackage{amsmath,epsfig}
\usepackage{amssymb,amsfonts}
\usepackage{latexsym}
\usepackage[latin1]{inputenc}
\usepackage{xcolor}

\usepackage{graphicx}
\usepackage{longtable}

\relax
\renewcommand{\theequation}{\arabic{section}.\arabic{equation}}

\def\be{\begin{equation}}
\def\ee{\end{equation}}
\def\ba{\begin{eqnarray}}
\def\ea{\end{eqnarray}}
\def\D{\Delta}
\def\t{\tau}
\def\a{\alpha}

\def\D{\Delta}

\def\g{\gamma}
\def\G{\Gamma}
\def\d{\delta}
\def\nn{\nonumber\\}
\def\pa{\partial}
\def\s{\sigma}

\title{Inverse Bootstrapping Conformal Field Theories}

\author{Wenliang Li\\
Department of Physics and Research Institute of Basic Science, Kyung Hee University, Seoul 02447, Korea}

\abstract{ 
We propose a novel approach to study conformal field theories (CFTs) in general dimensions. 
In the conformal bootstrap program, 
one usually searches for consistent CFT data 
that satisfy crossing symmetry. 
In the new method, we reverse the logic 
and interpret manifestly crossing-symmetric functions 
as generating functions of conformal data. 
Physical CFTs can be obtained by scanning the space of crossing-symmetric functions. 
By truncating the fusion rules, 
we are able to concentrate on the low-lying operators and derive  
some approximate relations for their conformal data. 
It turns out that the free scalar theory, the 2d minimal model CFTs, 
the $\phi^{4}$ Wilson-Fisher CFT, 
the Lee-Yang CFTs and the Ising CFTs 
are consistent with the universal relations from 
the minimal fusion rule $\phi_1\times \phi_1=I+\phi_2+T$, 
where $\phi_1,\,\phi_2$ are scalar operators, 
$I$ is the identity operator 
and $T$ is the stress tensor. 
}


\preprint{}

\begin{document}

\maketitle 

\section{Introduction}
Conformal field theories (CFTs) in general dimensions are quantum field theories 
that are invariant under global conformal transformations. 
They play an important role in various areas of theoretical physics, 
from critical phenomena to high energy physics. 

A CFT is characterized by correlation functions of its local operators. 
The spectral data of the local operators are their scaling dimensions and representations of the rotation group. 
The primary operators correspond to the lowest states of the dilation operator. 
By acting with the momentum operator, we obtain the descendant operators of higher scaling dimensions.  
Usually, a local operator is a linear combination of primaries and descendants. 
The Operator Product Expansion (OPE) of two local operators reads
\be
\mathcal O_i(x)\,\mathcal O_j(0)=\sum_k C'_{ijk}(x,\,\pa_y)\,\mathcal O_k(y),
\ee
where $k$ runs over the primary operators and $\pa_y^n \mathcal O_k(y)$ are the descendants. 
The power series $C'_{ijk}(x,\,\pa_y)$ in $\pa_y$ are determined by conformal invariance 
up to some multiplicative factors $C_{ijk}$ 
called OPE coefficients or structure constants of the operator algebra. 
\footnote{If there is more than one possible tensor structure, $C_{ijk}$ is a set of numbers. } 
The normalized two-point functions of the primary operators are fixed by conformal symmetry, 
while the three-point functions are determined by conformal invariance up to the OPE coefficients. 
By operator product expansions, 
n-point functions reduce to sums of $(n-1)$-point functions. 
Therefore, all the information of the correlation functions of the local operators is encoded in the spectral data and the OPE coefficients, which are called the CFT data. 

The OPE algebra is associative because correlation functions are independent of 
how the operator product expansions are performed. 
In the conformal bootstrap program \cite{Ferrara:1973yt,Polyakov:1974gs,Belavin:1984vu}, 
OPE associativity is promoted to a dynamical principle. 
This non-perturbative approach is not based on the Lagrangian or Hamiltonian formalism. 
Instead, one attempts to solve conformal field theories 
using only the consistency conditions from the OPE associativity, 
together with some physical assumptions, such as unitarity or fusion rules. 
In the case of four-point functions, the non-trivial consistency conditions for the CFT data 
are the equivalence of three possible OPE channels. 
They are also called crossing equations. 

In 2d, the global conformal symmetry is extended to the infinite dimensional Virasoro symmetry. 
Exact solutions can be obtained when the number of Virasoro primaries is finite. 
In particular, the exact solutions of the 2d minimal model CFTs \cite{Belavin:1984vu} 
are among the most important applications of the conformal bootstrap program. 
\\

In the modern numerical bootstrap approach \cite{Rattazzi:2008pe} 
(see \cite{Rychkov:2009ij}-\cite{Chang:2017xmr} for later developments), 
significant progress has been made in determining the low-lying CFT data of various CFTs. 
By studying the crossing equations geometrically, 
the possible solution regions are bounded by the positivity constraints from unitarity. 
As a prominent example, the 3d Ising CFT does not have an exact solution yet, 
but its low-lying CFT data can be determined to high precision, 
as the relevant parameter space of the lowest scalars is confined to a small isolated region  
by considering a set of four-point functions \cite{Kos:2014bka, Simmons-Duffin:2015qma, Kos:2016ysd}. 
Although unprecedented precision has been achieved,  
the CFT data of high spin and large scaling dimension operators remain unknown. 

It is important that the infinite number of subleading operators are irrelevant to the achieved precision. 
This indicates that the CFT data of these low-lying operators already 
provide a successful approximation of the 3d Ising CFT 
which is consistent with crossing symmetry. 
It reminds us of the spirit of Effective Field Theory, 
where the effective description is insensitive to 
the operators beyond the cutoff. 
In the context of AdS/CFT \cite{Maldacena:1997re,Gubser:1998bc,Witten:1998qj}, 
the effective description of a boundary CFT is related to 
the decoupling of heavy massive states in the bulk \cite{Fitzpatrick:2010zm}. 
\footnote{Note that we do not assume all CFTs have bulk dual theories.}

The success of the modern numerical bootstrap approach is a surprise 
\footnote{A plausible explanation is that the conformal block expansion 
converges exponentially fast in the Euclidean regime \cite{Pappadopulo:2012jk,Rychkov:2015lca}. 
}, 
as crossing symmetry is expected to relate one low-lying operator 
to an infinite number of fast spinning operators of large scaling dimensions. 
For example, in the analytic bootstrap approach, 
some general properties of the high spin spectrum can be deduced 
from the light-cone limits of the crossing equations \cite{Fitzpatrick:2012yx,Komargodski:2012ek}. 
\footnote{See \cite{Simmons-Duffin:2016wlq,Vos:2014pqa,Kaviraj:2015cxa,Alday:2015eya,Kaviraj:2015xsa,Alday:2015ota,Alday:2015ewa,Dey:2016zbg,Alday:2016mxe,Alday:2016njk,Alday:2016jfr} 
for more results along the lines of large spin expansions.} 
On the contrary, the developments in the numerical approaches indicate that  
crossing symmetry also gives surprisingly strong constraints on the low-lying CFT data themselves. 

To understand this surprise, 
we want to investigate two questions: 
\begin{itemize}
\item
How much are the low-lying CFT data constrained by crossing symmetry? 
\item
When does a truncated spectrum provide a consistent, effective description? 
\end{itemize}
These two questions are related and overlapping. 
On the one hand, the CFT data of the low-lying spectrum are heavily constrained by crossing symmetry  
when the crossing equations are approximately solved by the low-lying spectrum. 
On the other hand, if a truncated spectrum almost solves the crossing equations, 
then the CFT data of the operators above the cutoff can only induce small perturbations of the low-lying CFT data, 
and their OPE coefficients should be much smaller than those of the low-lying operators. 
Therefore, the essential point is whether the low-lying/truncated CFT data 
approximately solve the crossing equations. 
\footnote{The observation that 
the numerical unitary bounds are converging and almost saturated by 
the 2d minimal models and the 3d Ising CFT 
indicates the crossing equations are approximately solved by the truncated spectra of these physical CFTs, 
which is related to the extremal functional method \cite{Poland:2010wg,ElShowk:2012hu,El-Showk:2014dwa,Mazac:2016qev,Simmons-Duffin:2016wlq}.
}
\\

To solve the crossing equations, 
the standard way in the modern bootstrap approach  
is to expand the equations around the crossing symmetric point $u=v=1/4$. 
In this work, we will solve the crossing equations in a different manner. 
Instead of expanding the crossing equations, 
we will directly construct the crossing solutions using explicitly crossing-symmetric functions. 
We would like to call this approach ``inverse conformal bootstrap". 

To proceed further, we will propose a truncation ansatz. 
In this truncation framework, we can concentrate on a few low-lying operators
and derive some relations for the CFT data of the truncated spectrum. 
This truncation ansatz is in spirit analogous to the Shifman-Vainshtein-Zakharov sum rules in QCD \cite{Shifman:1978bx,Shifman:1978by}. 
In this work, we will focus on the minimal fusion rule of two identical scalar operators
\footnote{The crudest fusion rule is given by
\be
\phi_1\times \phi_1 = I+\phi_2.
\ee
In section \ref{toy-ex}, this truncated OPE will be discussed in detail 
as the simplest example of our truncation ansatz. 
}
\be
\phi_1\times \phi_1 = I+\phi_2+T, \label{mini-fr}
\ee
where $\phi_1,\,\phi_2$ are scalar primary operators, $I$ is the identity operator and $T$ is the stress tensor
\footnote{The idea of solving the crossing equations by a severely truncated fusion rule was first proposed by Gliozzi \cite{Gliozzi:2013ysa}. See \cite{Gliozzi:2014jsa,Gliozzi:2015qsa,Nakayama:2016cim,Gliozzi:2016cmg,Esterlis:2016psv} also for some later results.}. 
We assume there is one relevant operator in the original fusion rule and it is a scalar. 
\footnote{It is also possible that there is no relevant operator, which was examined in the 2d CFTs in \cite{Esterlis:2016psv}.}
We also consider longer fusion rules for 2d CFTs 
to capture the decoupling of subleading operators. 

Using the truncation ansatz, 
the OPE coefficients $P_2,\,P_T$ are approximated by some rational functions 
of the scaling dimensions $\D_1,\,\D_2$ and the spacetime dimension $d$.
Interestingly, it turns out many physical CFTs are consistent with the same equations, 
so these relations are universal! 
This phenomenon echoes the emergence of universality in low energy physics 
as the effective description is insensitive to the microscopic details.  
\\

The paper is organized as follows: 

In section \ref{inv-bootstrap}, 
we describe our inverse approach to the conformal bootstrap program and 
a natural truncation ansatz arising from this perspective. 
The general truncation procedure is illustrated by a toy example.

In section \ref{2d-CFT}, we apply the truncation framework to the conformal field theories in two dimensions. 
We focus on the 2d minimal models 
where the exact expressions of the 4-point functions are known, 
so we can compare our estimates of the OPE coefficients with the exact values. 
We also show how one can identify the 2d Lee-Yang CFT and the 2d Ising CFT in the truncation framework 
based on the phenomenon of operator decoupling. 

In section \ref{GenD-CFT}, 
we consider conformal field theories in general dimensions. 
In contrast to section \ref{2d-CFT}, we assume the presence of twist gaps. 
After deriving the approximate equations of the OPE coefficients, 
we examine several physical CFTs in various dimensions 
and show the universal equations are consistent with the well-established results. 

In section \ref{discuss}, 
we further discuss our results and propose some directions for future investigations. 

\section{Inverse bootstrapping method} \label{inv-bootstrap}
In the quantum inverse scattering method \cite{Takhtajan:1979iv}, 
one begins with the solutions of a non-trivial consistency condition, 
i.e. the Yang-Baxter equation \cite{Yang:1967bm,Baxter:1972hz}. 
Analogously, in the inverse bootstrapping method, 
we will start from the solutions of a non-trivial consistency condition, 
namely the crossing equation \eqref{crossing-1} to be defined below
\footnote{See also the recent works \cite{Gadde:2017sjg,Hogervorst:2017sfd,Hogervorst:2017kbj} 
which study the systematic constructions of crossing symmetric solutions.}. 
The CFT data can be directly deduced from a given crossing symmetric function.
By working at the level of the solution space, 
the inverse perspective provides us with 
a natural truncation ansatz 
obeying the crossing equation \eqref{crossing-1}. 

\subsection{Crossing symmetric functions}
In Polyakov's original paper on the conformal bootstrap \cite{Polyakov:1974gs}, 
he devised an alternative approach \footnote{This approach was revisited and extended in the recent works 
\cite{Sen:2015doa,Gopakumar:2016wkt,Gopakumar:2016cpb,Dey:2016mcs}. } 
which was less explored compared to the standard one.
In this approach, one starts from some explicitly crossing-symmetric building blocks 
which however contain unphysical logarithmic terms. 
Then physical operator product expansions require some consistency conditions 
so that the logarithmic terms cancel out, 
leading to constraints on the CFT data. 
As an example, Polyakov reproduced the lowest order anomalous dimensions 
of the Wilson-Fisher fixed point in $d=4-\epsilon$ dimensions. 

The idea of our inverse approach is similar to Polyakov's alternative method, 
but we use some information of the general structure of conformal blocks,  
so our crossing-symmetric building blocks are compatible with operator product expansions.
The starting point is the non-trivial crossing equation
\be
v^\D\, G(u,v)=u^\D\, G(v,u),\label{crossing-1}
\ee
where $G(u,v)$ is the conformal invariant part of the four-point function of identical scalar operators
\be
< \phi(x_1)\,\phi(x_2)\,\phi(x_3)\,\phi(x_4)>
=\frac 1 {(x_{12}^2 x_{34}^2)^{\D}} G(u,v),
\ee
the conformal invariant cross-ratios are defined as
\be
u=\frac {x_{12}^2 x_{34}^2}  {x_{13}^2 x_{24}^2}
,\quad
v=\frac {x_{14}^2 x_{23}^2}  {x_{13}^2 x_{24}^2},
\ee
and $\D$ is the scaling dimension of the external scalar operator $\phi$. 
\footnote{
There is another crossing equation
\be
G(u,v)=G(u/v,1/v).\label{crossing-2}
\ee
Since we only consider four-point functions of identical scalar operators, 
the second crossing equation \eqref{crossing-2} is solved if all the exchanged operators have even spins. 
We focus on the first crossing equation \eqref{crossing-1} in this work.}
\\ 

Let us introduce a two-variable function
\be
H(u,v)= v^\D\, G(u,v).
\ee
The crossing equation \eqref{crossing-1} indicates $H(u,v)$ is a symmetric function of $u$ and $v$. 

If $H(u,v)$ is a symmetric polynomial function, 
it can be decomposed into
\be
H(u,v)=\sum_{m,n=0}^\infty c^{m,n}\,(u^m v^n+u^n v^m),\quad m\leq n,
\ee
where the crossing symmetric building blocks are
\be
u^m v^n+u^n v^m.
\ee

For example, in 4d free CFTs, 
the 4-point function of the canonical scalar corresponds to 
\be
H_{\text{4d free}}(u,v)=v (1+u+u/v)= uv+ (u+v). 
\label{H-4d-free}
\ee
In general, $H(u,v)$ is not necessarily a polynomial function. 
For instance, in the 2d Ising CFT, the four point function of the spin operator corresponds to
\be
H_{\text{2d Ising}}(u,v)=\frac {\sqrt{1+\sqrt u+\sqrt v}}{\sqrt 2}. 
\ee
If we expand $H(u,v)$ around $u=v=0$, 
the crossing symmetric building blocks are
\be
u^m v^n+u^n v^m,\qquad
u^{1/2+m} v^n+u^n v^{1/2+m},
\ee
where the exponents are extended to rational numbers. 
The half-integer exponents $1/2+m$ have clear physical interpretation in the operator product expansions: 
they are related to the twists of the exchanged operators.
To see their physical interpretation, 
we need to examine the series expansion of conformal blocks 
(see Appendix \ref{App-series-cb} for more details), 
which reads
\be
F_{\t,\,l}(u,\,v)=u^{\t/2}(1-v)^l[1+\mathcal O(1-v)]+\mathcal O (u^{\t/2+1}),
\ee
where $\t,\,l$ are the twist and the spin of the exchanged primary operator. 
\footnote{The scaling dimension of the exchanged primary operator is
\be
\D=\t+l.
\ee
For scalar operators, the twists are equal to the scaling dimensions. 
In this work, we use twists as the independent spectral parameters, instead of scaling dimensions. 
} 
From the general structure of the series expansion, 
we know the exponents of the symmetric power functions are given by $\t/2+m$. 

In the 2d Ising CFT, the four-point function of the spin operator is 
a linear combination of two Virasoro conformal blocks, 
corresponding to the identity operator and the energy operator. 
We can decompose the Virasoro block of the energy operator into global blocks. 
One can check that the twists of the low spin primary operators are precisely $\t=1$, 
which explains the presence of 1/2 in the exponents.
Note that, in a generic CFT, the twists of primary operators are not restricted to rational numbers. 
Furthermore, in a non-unitary CFT, the twists can also violate the unitary bounds. 
\\

In (generalized) free CFTs and 2d Minimal models, 
where exact expressions of the 4-point functions are known, 
we can explicitly decompose $H(u,v)$ into the crossing symmetric building blocks
\be
u^{\t/2+m} v^{\t'/2+n}+u^{\t'/2+n} v^{\t/2+m},\label{sym-bb}
\ee
where $\t,\,\t'$ are the twists of the exchanged primary operators 
\footnote{When the external scalars are different, 
the crossing symmetric building blocks are not necessarily symmetric functions, 
as the exchanged operators in s- and t- channels may be different. }.

As a working assumption, we assume 
$H(u,v)$ can be decomposed into the crossing symmetric building blocks \eqref{sym-bb}. 
It not clear what are the necessary conditions 
for the existence of this series representation. 
This issue was also discussed in \cite{Alday:2015ota}. 

If we expand the symmetric function $H(u,v)$ by a smooth deformation parameter, 
up to some subtleties of potential infinite sums, 
we should obtain symmetric polynomials of $\log u,\,\log v$ at each order. 
In addition, the degrees of the polynomials should match with the expansion orders. 
In \cite{Alday:2015ota}, Alday and Zhiboedov conjectured that  
this is always the case in the four-point functions in weakly coupled CFTs, 
where the deformation parameters correspond to some small coupling constants. 
As test examples, 
they checked the four-point function of identical half BPS scalars in $\mathcal N=4$ supersymmetric Yang-Mills theory \cite{Eden:2000mv,Bianchi:2000hn} to two-loop order 
and the correlators of Konishi operators \cite{Bianchi:2001cm} to one-loop order.  

In the $\epsilon$-expansion, we have a different deformation parameter. 
In the $\phi^4$ Wilson-Fisher CFT, 
the parameter is $\epsilon=4-d$ 
and the zeroth order expression is given by \eqref{H-4d-free}. 
We expect and check that the first order correction should be a degree-1 polynomial in $\log u$ and $\log v$ 
\be
\Big[h_1(u,v)\log u+h_2(u,v)\log u \log v+(u\leftrightarrow v)\Big]\,\epsilon, 
\label{H-epsilon}
\ee
where the classic results in \cite{Wilson:1971dc} are used
\footnote{Note that $h_1(u,v)$ and $h_2(u,v)$ are power series of $u,v$. 
We only verified \eqref{H-epsilon} to a few orders. 
In principle, \eqref{H-epsilon} can be proved using the closed form expression of the scalar conformal block.}
\footnote{ 
A more non-trivial check would be the $\epsilon^2$ order terms 
which can be done using the second order correction to the OPE coefficients 
recently obtained in \cite{Gopakumar:2016wkt,Gopakumar:2016cpb}.}.

The Mellin representation of $H(u,v)$ is a contour integral of $u^s v^t\,\mathcal M(s,t)$, 
where $\mathcal M(s,t)$ is the Mellin amplitude up to some normalization convention. 
We expect in the Mellin space the crossing equation \eqref{crossing-1} translates into the requirement 
that poles are crossing-symmetric in s- and t- channels. 
A celebrated example of crossing symmetric pole structure is the Veneziano amplitude \cite{Veneziano:1968yb} 
involving the Gamma function. 
In 2d CFTs, one finds that the Mellin amplitudes 
\footnote{The use of Mellin transform in the context of CFTs was initiated in \cite{Mack:2009mi}.} of the 2d Ising CFT 
and the 2d Lee-Yang CFT are also given by products of Gamma functions 
with rational poles symmetric in the s- and t- channels \cite{Alday:2015ota}. 
Furthermore, a recent work \cite{Rastelli:2016nze} 
on the holographic 4-point functions of half BPS operators 
in Type-IIB supergravity on $AdS_5\times S^5$ 
shows that, in the case of identical operators, a simple Mellin amplitude with 
manifestly crossing-symmetric poles can reproduce the previous explicit results in 
\cite{Arutyunov:2000py,Arutyunov:2002fh,Arutyunov:2003ae,Dolan:2006ec}. 
The formulation of the bootstrap ansatz in \cite{Rastelli:2016nze} 
is similar to our inverse bootstrap perspective. 

In fact, in \cite{Dolan:2006ec}, which is based on the position space, 
the general 4-point functions are directly constructed from 
a finite number of crossing symmetric building blocks. 
In some sense, our inverse bootstrap ansatz (see the precise formulation in \eqref{H-cross-sym}) is 
a non-supersymmetric generalization of the ansatz in \cite{Dolan:2006ec}. 
To proceed further, we will introduce a truncation procedure 
so that the less symmetric CFTs become finite-dimensional problems as well. 
\\

The free correlators are clearly linear combinations of the crossing symmetric building blocks \eqref{sym-bb}. 
We argue that if some interacting CFTs are connected to free theories by smooth, continuous deformations, 
$H(u,v)$ should also allow an expansion in terms of the crossing symmetric building blocks \eqref{sym-bb}
\be
H(u,v)=\sum_{i,j}\sum_{m,n=0}^\infty c_{i,j}^{m,n}\,
(u^{\t_i/2+m} v^{\t_j/2+n}+u^{\t_j/2+n} v^{\t_i/2+m}),
\label{sym-sum-H}
\ee 
where $\t_i$ runs over the twist spectrum from low to high twists 
and the indices run over two possibilities
\be
i<j,
\quad
\text{or}\quad
i=j,\quad m\leq n. 
\ee
If $i=j$ and $m=n$, the summand has only one term to avoid over-counting. 

A concise, equivalent representation of \eqref{sym-sum-H} is
\be
H(u,v)=\sum_{i,j}\sum_{m,n=0}^\infty c_{i,j}^{m,n}\,
u^{\t_i/2+m} v^{\t_j/2+n},
\label{H-cross-sym}
\ee 
where the crossing equation \eqref{crossing-1} requires
\be
c_{i,j}^{m,n}=c_{j,i}^{n,m}.
\ee 

If the twist spectrum is discrete and bounded from below, 
then one can reconstruct the series order by order. 
Note that we do not assume unitarity. 

Let us first consider the limit $u\ll 1$ but $v$ fixed at a finite value. 
From the s-channel OPE expansion,  
we can identify the minimal twist $\t_1$ from the leading asymptotic behavior of $H(u,v)\sim u^{\t_1/2}f_1(v)$. 
The second lowest twist can be determined from the leading asymptotic behavior of 
$H(u,v)-u^{\t_1/2}f_1(v)\sim u^{\t_2/2}f_2(v)$. 
In principle, we can recover the twist spectrum by repeating this procedure. 
Note some of them are the twists of the descendants. 

From crossing symmetry, this series expansion is dominated 
by crossing symmetric building blocks \eqref{sym-bb} of small exponents. 
Then we can consider the limits $u\ll1,\, v\ll 1$ to fix 
the coefficients $c_{i,j}^{m,n}$ of the crossing symmetric building blocks \eqref{sym-bb}. 
For example, we have $f_1(v)\sim c_{1,1}^{0,0}\, v^{\t_1/2}$. 
In principle, we can reconstruct the series expansion \eqref{H-cross-sym} order by order 
in the small $u,v$ expansion. 

Naively, in the lightcone limit, where $u,v\rightarrow 0$, 
the leading crossing-symmetric building block is $u^{\t_\text{min}/2}\,v^{\t_\text{min}/2}$. 
For instance, we have $\t_\text{min}=0$ in the 2d Ising CFT, while $\t_\text{min}=-2/5$ in the 2d Lee-Yang CFT. 
However, this leading term can be absent. 
If the identity has the lowest twist and the twist spectrum is gapped, 
the leading crossing-symmetric combination becomes $u^{\t_\text{min}/2}\,v^\D+u^\D\,v^{\t_\text{min}/2}$ with $\t_\text{min}=0$. 
\\

Now let us introduce the concept of ``twist family" 
\footnote{See \cite{Alday:2016njk,Alday:2016jfr} also for the recently proposed interesting concept ``twist conformal block", 
which is a special combination of global conformal blocks in the same twist family. }
\be
\{\t_i\}:=\frac{\t_i}2,\,\frac{\t_i}2+1,\,\frac{\t_i}2+2,\,\frac{\t_i}2+3,\,\dots,
\ee
which generalizes the Virasoro modules in 2d CFTs and will be useful later. 
In $H(u,v)$, the exponents in the form $\t_i/2+m$ are in the same twist family $\{\t_i\}$. 
They are related to exchanged operators of twists $\t_i+2m$. 
\footnote{The minimum value of the exponents in a twist family $\{\t_i\}$ can be larger 
than $\t_i/2$ if there is more than one exchanged operator with twist $\t_i$ 
and the lowest order term cancels out. } 
In the s-channel OPE, we can identify the contributions of the exchanged operators 
in the same twist family with a partial sum by fixing $i$
\be
\sum_{\t_{\mathcal O}\subset\{\t_i\}} P_{\mathcal O}\, F_{\t_{\mathcal O},\,l_{\mathcal O}}(u,v)
=\sum_{m,n=0}^\infty u^{\t_i/2+m} v^n
\sum_{j} c_{i,j}^{m,n}\, v^{\t_j/2-\D},
\ee
where $F_{\t_{\mathcal O},\,l_{\mathcal O}}(u,v)$ is the conformal block of the exchanged operator $\mathcal O$. 
These partial sums generalize the Virasoro conformal blocks in 2d CFTs. 

If there is a finite number of exchanged operators in the same twist family $\{\t_i\}$, 
then the left hand side contains logarithmic terms
\be
u^{\t_i/2+m}\, v^n \log v
\ee
from the small $v$ asymptotic behavior of the conformal blocks. 
In order to reproduce the same logarithmic singularity, 
the right hand should contain power law singularities due to negative exponents
\be
\t_j/2-\D<0.
\ee
Then we can expand the power functions
\be
v^{\t_j/2-\D}=e^{(\t_j/2-\D)\log v}=1+(\t_j/2-\D)\log v +\mathcal O[(\log v)^2], 
\ee
which requires an infinite number of negative exponent terms  
to cancel all the higher order logarithmic singularities. 
In particular, in unitary CFTs where the primaries obey the unitary bounds,  
there exist an infinite number of exchanged spinning operators $\mathcal O_i$
whose anomalous dimensions satisfy
\be
0\leq \g_i < 2\g_{\phi},
\ee
where $\g_{\phi}=\D-(d-2)/2$ is the anomalous dimension of the external scalar. 
We have assumed only a finite number of exchanged scalar operators have scaling dimensions smaller than $2\D$.
In this way, we deduce a general property of the high spin spectrum 
first obtained in \cite{Fitzpatrick:2012yx,Komargodski:2012ek}.

\subsection{Truncating the fusion rules}
For each symmetric function in the form of \eqref{H-cross-sym},
we can deduce the corresponding spectrum and the OPE coefficients of the exchanged operator
\be
G(u,v)=v^{-\D}H(u,v)=1+\sum_{k} P_k\, F_{\t_k,\,l_k}(u,v).
\ee
where $P_k$, to be more precise, are the squares of the OPE coefficients 
\be
P_k=C^2_{\phi\phi\mathcal O_k}.
\ee 

In principle, we can search for physical CFTs by scanning the space of crossing symmetric functions. 
However, to match with the conformal partial wave expansions, we need to expand $v$ around $v=1$
\be
v^{\a}=[1-(1-v)]^\a=1-\a(1-v)+\mathcal O[(1-v)^2].
\ee 
Then the high order terms also contribute to the low-lying conformal data. 
We need to know the exact expression of $H(u,v)$ to obtain the exact CFT data of the low-lying operators. 

In a generalized free CFT, 
$H^{\text{free}}(u,v)$ involves only one crossing symmetric building block and 
the free parameter is the scaling dimension of the external scalar operator. 
But in a generic interacting CFT, we have an infinite number of parameters. 
In some 2d CFTs, 
we can group the global CFT data according to 
a finite number of Virasoro primary operators, 
then the parameter spaces have finite dimensions. 
In general, the parameter spaces of interacting CFTs seem to be infinitely dimensional. 
\\

At the level of crossing symmetric functions, we can impose functional truncations. 
In other words, 
we can reduce the dimension of the parameter space to a finite number 
by restricting to the subspace of crossing symmetric functions 
which are constructed from 
a finite number of crossing symmetric building blocks \eqref{sym-bb}. 
In this inverse approach, we are equipped with natural truncation schemes 
which are compatible with the crossing equation \eqref{crossing-1}. 

In this work, the truncation ansatz consists of three steps:
\begin{itemize}
\item
Step 1

In the first step, we truncate the physical fusion rule to a few low-lying operators
\be
\phi\times\phi=I+\mathcal O_1+\mathcal O_2+\dots+\mathcal O_p,
\label{fusion-trun}
\ee 
so we have a finite number of exchanged primary operators. 
Here, we impose a cutoff and omit the primary operators of small OPE coefficients. 
The exponents of the crossing symmetric building blocks are grouped into a finite number of sets 
according to their twist families. 

\item
Step 2

Although we have truncated the fusion rules, 
$H(u,v)$ still involves an infinite number of exponents. 
In the second step, we introduce a cutoff $M$ for the twists. 
In each twist family $\{\t_i\}_M$, the possible exponents are then limited to 
\be
\{\t_i\}_M\rightarrow \frac{\t_i}2,\,\frac{\t_i+2}2,\,\dots,\,\frac{\t_i+2M-2}2,\,\frac{\t_i+2M}2.
\ee
Now we have only a finite number of crossing symmetric building blocks 
from the truncated twist families $\{\t_i\}_M$. 
Let us denote the number of inequivalent twist families by $N$
\be
\{\t_i\}_M,\quad \text{with}\quad i=1,2,\dots, N. 
\ee 

\item
Step 3

In the third step, 
we require $G(u,v)=v^{-\D}\,H(u,v)$ reproduces the truncated fusion rule \eqref{fusion-trun}. 
This amounts to expanding $G(u,v)$ around $\{u,\,v\}=\{0,\,1\}$ 
and matching the series coefficients with those of the conformal blocks. 
For each twist family, we impose
\be
u^{\t_i/2} \sum_{m,n=0}^M \sum_{j=1}^N c_{i,j}^{m,n}\, 
u^{m} 
\,v^{\t_j/2+n-\D}
\sim
\sum_{\t_{\mathcal O}\subset\{\t_i\}} P_{\mathcal O}\, F_{\t_{\mathcal O},l_{\mathcal O}}(u,v),
\label{match}
\ee
where $M$ is the twist cutoff and $N$ is the number of twist families. 
Now we introduce the descendant cutoff $K$ and 
the matching is valid for the low lying spectrum of the twist family $\{\t_i\}_M$
\be
\D\leq\t_i+K,\quad \t\leq\t_i+2M,
\ee
where the second equation is from Step 2. 
The descendant cutoff $K$ can also be understood as a scaling dimension cutoff. 

Intuitively, we can interpret the scalar primary with scaling dimension $\t_i$ 
as the higher dimensional generalization of the Virasoro scalar primary. 
Then a term
\be
u^{\t_i/2+m}(1-v)^l
\ee
is associated with a level $k=2m+l$ descendant with spin $l$ and scaling dimension $\D_{i,k}=\t_i+2m+l$.

To exactly reproduce the truncated fusion rule, 
we need an infinite number of crossing-symmetric building blocks. 
Since we introduce a cutoff $M$ to the maximum twist in the second step, 
the left hand side of \eqref{match} already miss some high-twist terms. 
In addition, even if we focus on the low-twist terms, 
the fusion rule from $H(u,v)$ will inevitably involve additional operators with large scaling dimensions. 

In technical terms, 
when we expand the left hand side of \eqref{match} around $v=1$, 
only the low order terms can match with the right hand side, 
because the parameter space is now finite-dimensional. 
Beyond the descendant cutoff $K$, the series coefficients will not match, 
which translates into the presence of additional operators with dimensions $\D> \tau_i+K$. 
The operators beyond the descendant cutoff may be related to 
physical conformal multiplets with large scaling dimensions 
in the untruncated spectrum.

The precise matching conditions can be expressed as
\ba
&&u^{\t_i/2} \sum_{j=0}^N \sum_{m,n=0}^M c_{i,j}^{m,n}\, 
u^{m} 
\,v^{\t_j/2+n-\D}
-
\sum_{\t_{\mathcal O}\subset\{\t_i\}} P_{\mathcal O}\, F_{\t_{\mathcal O},l_{\mathcal O}}(u,v)
\nn
&=&\sum_{m=0}^{M}u^{\t_i/2+m}\, \mathcal O[(1-v)^{max(K-2m+1,\,0)}]
+\mathcal O(u^{\t_i/2+M+1}),
\label{mat-1}
\ea
where, in the second line, 
the first term is due to the descendant cutoff $K$ and 
the second term is related to the twist cutoff $M$. 
We have used some information of the series expansion of spinning conformal blocks. 
We also use the $max$ function to avoid negative exponents. 
\end{itemize}

Above, we explain the general truncation procedure. 
Let us emphasize that we are not considering the lightcone expansion around $u=v=0$. 
In the truncated functional space, the ``effective" coefficients $c_{i,j}^{m,n}$ of the same CFT 
depend on the truncation cutoffs 
\footnote{Although $c_{i,j}^{m,n}$ do not converge in the severe trunctions, 
the low-lying CFT data usually become more accurate as we introduce more primary operators and increase the twist and descendant cutoffs.} 
and they are different from those in the lightcone expansion 
\footnote{The lightcone expansion may correspond to the limit where the cutoffs are sent to infinity, 
in analogy to the UV fixed point of a renormalization group flow.}. 
The power function building blocks \eqref{sym-bb} are just an intuitive and efficient basis for the crossing-symmetric functions 
\footnote{As we will see in section \ref{GenD-CFT}, the results of the 2d CFTs are reasonably accurate 
even if we use a ``wrong" basis from the perspective of the lightcone expansion. 
The physical CFT data are not very sensitive to the difference in the two bases. 
The exceptions are the OPE coefficients of the stress tensor in the 2d Lee-Yang and Ising CFTs, 
where the first order estimate \eqref{2d-PT-1} generates the exact values. }. 
One should think of $v^{-\D}H(u,v)$ as a trial crossing solution expanded around $\{u,\,v\}=\{0,\,1\}$. 
\footnote{A countable basis for the crossing-symmetric functions is 
\be
(u-a)^m (v-a)^n+(u-a)^n (v-a)^m,\quad m,n=0,1,2,\dots\,,
\ee
around the Euclidean crossing symmetric point $u=v=a\geq 1/4$.}
The OPE convergence is rapid as the expansions are in the Euclidean regime
\footnote{The Euclidean regime is defined by $4u\geq(u-v+1)^2$.}.

The assumption in the truncation procedure is that the OPE coefficients of the subleading operators are 
much smaller than those of the low-lying operators. 
The low-lying operators are those with low twists $\t$ and small scaling dimensions $\D$. 
\footnote{Since the spin $l$ of an operator is the difference between its scaling dimension $\D$ and twist $\t$, i.e. $l=\D-\t$. 
The low-lying operators have low spins as well. 
We assume $\D$ and $\t$ are bounded from below. }
Step 1 is about primary operators and the corresponding conformal multiplets, 
while Step 2 and Step 3 are about descendant operators. 
The physical CFTs should be the case due to the rapid OPE convergence. 

From the matching conditions \eqref{mat-1}, 
we obtain a set of polynomial equations
\be
\frac{1}{k!}
\sum_{j=0}^N\sum_{n=0}^M 
(\D-\t_j/2-n)_k\,
c_{i,j}^{m,n}=
\sum_{\t_{\mathcal O}\subset\{\t_i\}} P_{\mathcal O}\, b_{m,k}(\t_{\mathcal O},\,l_{\mathcal O},\,d),
\ee
\be
i=0,1,\dots,N,\quad
k=0,1,\dots, K-2m,
\ee
where $b_{m,k}(\t_{\mathcal O},l_{\mathcal O},d)$ 
are the series coefficients of the conformal block of 
the exchanged operator $\mathcal O$ in d-dimensional spacetime 
(see Appendix \ref{App-series-cb}) 
and $(x)_n$ is the Pochhammer symbol
\be
(x)_n=\G(x+n)/\G(x).
\ee

We interpret these equations as a system of linear equations, 
where the unknowns correspond to the OPE coefficients $P_\mathcal O$ and 
the crossing symmetric function coefficients $c_{i,j}^{m,n}$. 
The coefficients of the linear system are polynomials of 
the scaling dimension $\D$ of the external scalar, 
the spectral data $\{\t_\mathcal O,\, l_\mathcal O\}$ and the spacetime dimension $d$. 
The linear system is non-homogeneous 
due to the fixed OPE coefficient of the identity operator. 

After choosing the truncated fusion rule and the twist cutoff $M$, 
the number of unknowns is fixed, 
so we can consider an appropriate descendant cutoff $K$ 
which lead to enough equations to solve the unknowns.  
\footnote{There might be several choices 
if the number of equations are slightly larger or smaller than that of the unknowns. } 
If the determinant is non-zero, we can solve the linear system. 
The solutions are rational functions. 
In the fractions, the denominators are given by the same determinant 
and the numerators are linear combinations of the determinants of the minors. 

In particular, the OPE coefficients are approximated by some rational functions
\footnote{A rational function is a fraction of two polynomial functions.} 
of the spectral data $\{\D,\,\t_\mathcal O,\, l_\mathcal O\}$ and the spacetime dimension $d$. 
Since we do not use any information of a specific CFT, 
the relations \footnote{Note that they are approximate relations because of the truncation procedure.} 
between the OPE coefficients and the spectral data 
are universal and model-independent! 
In other words, these relations should apply to different CFTs, 
as long as they are consistent with the truncated fusion rule. 

In section \ref{2d-CFT} and section \ref{GenD-CFT}, 
we examine some concrete CFTs in various dimensions. 
It turns out that many physical CFTs are consistent with the minimal fusion rule
\be
\phi_1\times\phi_1=I+\phi_2+T,
\ee
where $\phi_1,\,\phi_2$ are scalar operators, 
$I$ is the identity operator and $T$ is the stress tensor.  
\\

Let us discuss a subtlety concerning the degeneracy of twist-0 operators.  
The fusion rules of two identical scalars always contain the identity operator. 
In $d>2$, typically only the identity operator has a vanishing twist
\be
\t_{I}=0.
\ee
In unitary CFTs, when $d>2$, the twist spectra are gapped 
because of the unitary bounds
\be
   \D\geq
\begin{cases}
    (d-2)/2,& \text{if}\quad l=0;\\
    d-2+l,              & \text{if}\quad l>0,
\end{cases}
\ee
which translate into twist gaps between the identity operator and the exchanged operators
\be
   \t\geq
\begin{cases}
    (d-2)/2,& \text{if}\quad l=0;\\
    d-2,      & \text{if}\quad l>0.
\end{cases}
\ee
In some non-unitary CFTs, such as the 3d Lee-Yang CFTs, 
the twist-0 state is also non-degenerate in the twist spectrum. 

If the identity operator is the only operator with zero twist, 
we can deduce from the crossing equation \eqref{crossing-1} that 
$G(u,v)$ and $H(u,v)$ contain some universal terms
\be
G(u,v)=1+\left (\frac u v\right)^\D+\dots,
\quad
H(u,v)=v^\D+u^\D+\dots,
\ee
where the second terms correspond to operators in the twist family $\{\t=2\D\}$. 
In (generalized) free CFTs, all the exchanged operators are in the double-twist family. 
However, in interacting CFTs, we usually do not find operators with twists $2\D+2n$. 
Therefore, in the crossing-symmetric function $H(u,v)$, 
we will introduce a twist family $\{\t=2\D\}$, 
but impose the OPE coefficients in this double-twist family vanish 
when the descendant levels are lower than the cutoff $K$. 

In 2d CFTs, both the identity operator, the stress tensor and many higher spin operators have zero twist, 
so it is possible that $G(u,v),\,H(u,v)$ do not contain the universal terms. 
We will not introduce the double-twist family $\{\t=2\D\}$ in the study of 2d CFTs in section \ref{2d-CFT}. 
\\

Before moving to a toy example of the truncation ansatz, 
we want to compare our truncation ansatz with Gliozzi's ansatz \cite{Gliozzi:2013ysa}: 
\begin{itemize}
\item
In both methods, the fusion rules are severely truncated. 
The use of fusion rule truncations in our approach is inspired by Gliozzi's work \cite{Gliozzi:2013ysa}. 
Our motivation is to better understand the effective descriptions of CFTs 
and their general structure in the spirit of Effective Field Theory. 
To our understanding, 
Gliozzi's main motivation is to extend the standard numerical bootstrap method beyond unitary CFTs 
using the determinant technique. 
Our method also applies to non-unitary CFTs. 
\item
In our inverse approach, the crossing equation \eqref{crossing-1} is exactly solved by crossing symmetric functions. 
In Gliozzi's approach, 
the crossing equation \eqref{crossing-1} is approximately solved 
following the standard numerical approach \cite{Rattazzi:2008pe}. 
In more details, our approach expands the crossing solutions 
around the conformal block boundary condition point $(u,v)=(0,1)$, 
while Gliozzi's approach expands the crossing equation \eqref{crossing-1} 
around the crossing symmetric point $u=v=1/4$. 
As a result, our method involves polynomial equations, 
while Gliozzi's method deals with transcendental equations. 

In addition, our approach requires many additional unknowns to parametrize the crossing symmetric function $H(u,v)$. 
A related consequence is that 
we can consider even more severe truncations of the fusion rules than Gliozzi's method. 
But the number of unknowns grows much faster 
when we consider more operators.
\end{itemize}

\subsection{Toy example: $\phi_1\times \phi_1 = I+\phi_2$}\label{toy-ex}
Let us consider a toy example of the truncation ansatz. 
The truncated fusion rule reads
\be
\phi_1\times \phi_1 = I+\phi_2.\label{fr-ex}
\ee

The absence of the stress tensor may seem unphysical
\footnote{In generalized free CFTs, the stress tensor is usually absent from the OPEs. 
}, 
but this truncation can be consistent when $\phi_2$ is a relevant operator, i.e. $\D_2<\D_T=d$. 
The truncation ansatz outlined above can be carried out explicitly. 
The result is simple but non-trivial. 

We follow the three-step ansatz:
\begin{enumerate}
\item
To begin with, the conformal invariant part of $<\phi_1(x_1)\phi_1(x_2)\phi_1(x_3)\phi_1(x_4)>$ reads
\be
G(u,v)=v^{-\D_1}\sum_{i,j=0,2}\,\sum_{m,n=0}^\infty c_{i,j}^{m,n} u^{\t_i/2+m} v^{\t_j/2+n},
\quad
(i,m)\leq (j,n),
\ee
where the possible values of $\t_i$ are determined by the truncated fusion rule
\be
\t_0=0,\quad \t_2=\D_2. 
\ee 
Here we use the 2d scheme without taking into account the issue of twist gaps. 

The crossing equation \eqref{crossing-1} implies that $v^{\D_1} G(u,v)$ should be a symmetric function in $(u,v)$, 
so the coefficients $c_{i,j}^{m,n}$ satisfy the crossing symmetric conditions
\be
c_{i,j}^{m,n} 
=c_{j,i}^{n,m}. 
\ee

\item
In the second step, we set the twist cutoff to zero, namely
\be
M=0.
\ee
In other words, we truncate the crossing symmetric function to the lowest order in $u$, 
so $G(u,v)$ has only 4 terms
\be
G(u,v)=c_{0,0}^{0,0}\,v^{-\D_1}
+c_{0,2}^{0,0}\, v^{\D_2/2-\D_1}
+c_{0,2}^{0,0}\, u^{\D_2/2}v^{-\D_1}
+c_{2,2}^{0,0}\, u^{\D_2/2} v^{\D_2/2-\D_1}
,
\ee
where we have used the crossing symmetric condition 
\be
c_{2,0}^{0,0}=c_{0,2}^{0,0}. 
\ee

Let us introduce a simplified notation for this toy example
\be
c_{i,j} =c_{i,j}^{0,0}.
\ee

\item
In the last step, let us set the descendant cutoff to be
\be
K=1,
\ee
then the number of equations is the same as the number of unknowns.

According to the descendant cutoff, we expand the crossing symmetric function around $v=1$ to the first order
\be
G(u,v)=a_{0,0}+a_{0,1}(1-v)
+u^{\D_2/2}
\left[
a_{2,0}+a_{2,1}(1-v)
\right]
+\mathcal O[(1-v)^2]
,
\ee
where $a_{i,j}$ are related to $c_{i,j}$ by
\ba
a_{0,0}&=&c_{0,0}+c_{0,2},
\nn
a_{0,1}&=&c_{0,0}\,\D_1
+c_{0,2}\, (\D_1-\D_2/2),
\nn
a_{2,0}&=&c_{0,2}+c_{2,2},
\nn
a_{2,1}&=&c_{0,2}\,\D_1
+c_{2,2}\, (\D_1-\D_2/2).
\ea

Then we require that $G(u,v)$ reproduces the series expansions of the conformal blocks at low orders
\ba
G(u,v)&=&1+P_2 F_{\D=\D_2,l=0}(u,v)+\mathcal O[(1-v)^2]
\nn\quad
&=&1+P_2 \left[1+ b_{0,1}(\D_2,0,d) (1-v)\right]+\mathcal O[(1-v)^2],
\ea
where from the series expansion of conformal blocks (see Appendix \ref{App-series-cb}) we have
\be
b_{0,1}(\D_2,0,d)=\frac {\D_2} 4 . 
\ee

There are 4 equations for 4 unknowns $\{P_2, \,c_{0,0}, \,c_{0,2}, \,c_{2,2}\}$. 
The linear system is given by
\be
a_{0,0}=1,\quad a_{0,1}=0,\quad 
a_{2,0}=P_2,\quad a_{2,1}=P_2\, \frac {\D_2} 4.
\label{toy-eqs}
\ee

When the determinant is non-zero, i.e. $3\D_2 \neq 4\D_1$, the linear system \eqref{toy-eqs} has a unique solution
\be
c_{0,0}=1-\frac {2\D_1}{\D_2},\quad
c_{0,2}=\frac {2\D_1}{\D_2},\quad
c_{2,2}=\frac {2\D_1}{\D_2}\frac {4\D_1-\D_2}{3\D_2-4\D_1},
\ee
and
\be
P_2=\frac {4\D_1}{3\D_2-4\D_1}.
\label{P-ex}
\ee
\end{enumerate} 

The spacetime dimension $d$ does not play a role 
because the series coefficients of a conformal block are $d$-independent 
at the lowest twist order ($m=0$).  
The OPE coefficient $P_2$ is a simple rational function \eqref{P-ex} of the scaling dimensions $\D_1,\D_2$ 
of the scalar operators
\footnote{One may derive an analogous relation by directly expanding $v^{\D_1}[1+P_2\, G_{\D_2,0}(u,v)]$ 
around the crossing symmetric point $u=v=1/4$ 
and requiring the coefficients of $(u-1/4)$ and $(v-1/4)$ match. 
The result is a transcendental equation
\be
P_2=-\frac { \D_1}{(\D_1+v\,\pa_v-v\,\pa_u) F_{\D_2,0}(u,v)}\Big|_{u\rightarrow 1/4,\,v\rightarrow 1/4}.
\ee 
However, this estimate equation is not as accurate as \eqref{P-ex}.}.

If we use the scheme for $d>2$ CFTs, 
the identity part in the crossing symmetric functions will be replaced by the double-twist power function with $\t_0=2\D_1$. 
In addition, in the third step, the OPE coefficients for the double-twist block should vanish. 
With the same cutoff parameters $(M,K)=(0,1)$, 
we obtain a linear system and the solutions of $c_{i,j}$ are different. 
However, the solution of the OPE coefficient $P_2$ is the same as \eqref{P-ex}. 
The relation \eqref{P-ex} is independent of which scheme is used.

The degenerate case $\D_2  =4\D_1/3$ has a curious interpretation in 2d. 
In \cite{Esterlis:2016psv}, the same relation for the scaling dimensions was found numerically using Gliozzi's determinant method. 
Then it was identified with the exact solution of a special Virasoro fusion rule $\phi_1\times\phi_1=\phi_2$, where the identity operator is absent. 
When $\D_1$ is positive, the central charge is larger than one because $c=1+16\D_1$. 
The decoupling of the identity operator is captured by the degeneracy condition in the toy example. 
\\

From the severely truncated fusion rule \eqref{fr-ex}, 
we obtain a simple relation \eqref{P-ex} for the conformal data $P_2,\D_1,\D_2$. 
Given the scaling dimensions of $\phi_1,\phi_2$, 
we can estimate the OPE coefficient $P_2$ 
using the model-independent, d-independent equation \eqref{P-ex}.
In Table \ref{tab-ex}, we compare some estimates from \eqref{P-ex} with the known results. 
Although the estimate equation \eqref{fr-ex} is very simple, 
the first significant figures agree with these known values. 
In the case of (generalized) free CFTs, our estimate coincides with the exact value. 
In particular, the 3d estimates are surprisingly accurate. 
\\

\begin{table}[h!]
\centering
 \begin{tabular}{||c ||c c c c c c||} 
 \hline
 	$P_2$				& free  	& $\phi^4$ WF		& 2d LY 	&  2d Ising	& 3d LY  	& 3d Ising \\ [0.5ex] 
 \hline
 \hline
estimate 	& 2 		& $2-\epsilon$ 		& 	-4	&	0.2		& 	-4	& 	1.0	 \\
\hline
exact/numerical 			& 2 		& $2-2\epsilon/3$	&    -3.7 	&	0.25		&    -3.9	&	1.1\\ 
 \hline
\end{tabular}
\caption{The estimates of the OPE coefficient $P_2$ from \eqref{P-ex} 
and the known results from exact solutions \cite{Wilson:1971dc,DiFrancesco:1997nk} or numerical conformal bootstrap  \cite{Simmons-Duffin:2016wlq,Gliozzi:2013ysa}
in several CFTs. 
We only show two significant figures. 
Input parameters are the scaling dimensions of $\phi_1,\phi_2$ 
from the exact results \cite{Wilson:1971dc,DiFrancesco:1997nk} or numerical conformal bootstrap \cite{Simmons-Duffin:2016wlq,Gliozzi:2013ysa}. 
LY and WF stand for Lee-Yang and Wilson-Fisher. 
To derive the estimates of $P_2$ in the Lee-Yang CFTs, 
we only use the information that $\D_1=\D_2$ and the analytic estimate is $P_2=-4$. 
The relative errors in 3d interacting CFTs are smaller than those in 2d. }
\label{tab-ex}
\end{table}

If we increase the twist cutoff and the corresponding descendant cutoff, 
the estimates of $P_2$ become less accurate, 
which indicates some subleading operators should be taken into account. 

\section{2d minimal model CFTs}\label{2d-CFT}
In two dimensions, the global conformal symmetry is extended to the infinite dimensional Virasoro symmetry. 
Exact solutions of some CFTs are known 
as their operator algebras become finitely dimensional. 
In this section, we examine the 2d minimal models using our truncation ansatz. 

In the 2d minimal models \cite{DiFrancesco:1997nk}, 
the Virasoro primary operators $\phi_{r,s}$ are labeled by two integers $(r,s)$. 
In general, a null state appears at level $l=rs$ 
in the Verma module $V(c,\, \D_{r,s})$, 
where $\D_{r,s}$ are the scaling dimensions of the Virasoro primary operators $\phi_{r,s}$. 
Let us consider $\phi_{1,2}$ which has a null descendant at level 2. 
The 4-point function of $\phi_{1,2}$ satisfies a second order differential equation 
due to the level-2 null state. 
The general solution is a linear combination of two independent solutions. 
They correspond to two Virasoro conformal blocks from the Virasoro fusion rule
\be
\phi_{1,2}\times \phi_{1,2}
=\phi_{1,1}+\phi_{1,3}\,,
\ee 
where $\phi_{1,1}$ is the identity operator. 
The scaling dimensions of $\phi_{1,2},\,\phi_{1,3}$ are
\be
\D_{1,2}=\D,\quad \D_{1,3}=\frac 2 3 (1+4\D).
\ee

A Virasoro conformal block can be decomposed into 
an infinite number of global conformal blocks in the same twist family. 
In particular, the Virasoro conformal block of the identity operator 
contains the stress tensor block, 
as the stress tensor is a Virasoro descendant of the identity operator. 

There are two integration constants in the general solution of the 4-point function of $\phi_{1,2}$. 
One of them is fixed by the OPE coefficient of the identity operator. 
The ratio of the two constants is determined by the crossing equation \eqref{crossing-1}. 
The explicit solution reads
\be
G(z,\,\bar z)= F_{1,1}^{\text{Virasoro}}(z)\,F_{1,1}^{\text{Virasoro}}(\bar z)
+ P_{1,3}\, F_{1,3}^{\text{Virasoro}}(z)\,F_{1,3}^{\text{Virasoro}}(\bar z),
\label{G-2d-m}
\ee
where the (anti-)holomorphic Virasoro blocks are 
\ba
F_{1,1}^{\text{Virasoro}}(z)&=&(1-z)^{-\D} 
{}_2 F_1\left[\frac {1-2\D} 3,\,-2\D,\,\frac {2(1-2\D)}3;\, z\right],
\nn
F_{1,3}^{\text{Virasoro}}(z)&=&(1-z)^{\frac{1+\D}3} z^{\frac{1+4\D}3} 
{}_2 F_1\left[\frac {2(1+\D)}3,\,1+2\D,\, \frac{4(1+\D)}3;\, z\right],
\ea
the Virasoro OPE coefficient $P_{1,3}$ is
\be
P_{1,3}=\frac{
2^{1 - \frac 8 3(1 + \D)}\,
  \G\left(\frac {2(1 - 2 \D)}3\right)^2 \G(1 + 2 \D)^2 
  \left[ 
   \sin\left(\frac{\pi(1 + 16 \D)}{6}\right)
   -\cos\left( \frac{\pi(1 + 4 \D)} 3\right)\right]}
   {\pi\, \Gamma\left(\frac {7+4 \D}6\right)^2},
   \label{P13}
\ee
and the variables $z,\,\bar z$ are related to $u,v$ by 
\be
u=z \bar z,\quad v=(1-z)(1-\bar z). 
\ee
In the crossing symmetric solution \eqref{G-2d-m}, we have a free parameter $\D$. 
In unitary minimal models, it takes some positive rational values
\be
\D=\D_{1,2}=\frac 1 2-\frac 3{2(m+1)},
\quad m=3,\,4,\,5,\,\dots\quad 
\ee
and the central charge reads
\be
c=1-\frac 6{m(m+1)}.
\ee

A more general way to parameterize $\D$ is
\be
\D=\D_{1,2}=\frac 1 8
\left[
5-c-\sqrt{(1-c)(25-c)}
\right].
\ee
By extending the central charge $c$ to a continuous parameter, 
we obtain an interpolating solution between unitary minimal models \cite{Liendo:2012hy}. 
Non-unitary minimal models are also included 
where $\D$ are given by some rational numbers different from the unitary values. 
\\

Let us decompose $G(u,v)$ into global conformal blocks
\be
G=1+P_2\, G_{\t_2,\,l_2}+P_T\, G_{\t_T,\,l_T}
+P_3\, G_{\t_3,\,l_3}
+\dots
\ee
and the global fusion rule corresponds to
\be
\phi_1\times \phi_1=I+\phi_2+T+\phi_3+\dots\quad.
\ee
The global spectral data of the low-lying primary operators in the OPE are
\be
\D_1=\D_{1,2}=\D,\quad
\D_2=\D_{1,3}=\frac 2 3 (1+4\D),
\label{spectral-2d-m-1}
\ee
\be
\{\t_2,\,l_2\}=\{\D_2,\,0\},\quad
\{\t_T,\,l_T\}=\{0,\,2\},
\label{spectral-2d-m-2}
\ee
\be
\{\t_3,\,l_3\}=\{\D_2,\,2\}.
\ee
Note that the identity operator $I$ and the stress tensor $T$ have the same twist, 
while the operators $\phi_2,\,\phi_3$ also have the same twist. 

The global OPE coefficients of the low-lying operators are
\be
P_2=P_{1,3},\quad
P_T=\frac {\D(1+\D)}{2(5-4\D)},
\label{OPE-2d-m}
\ee
\be
P_3=-\frac {(1+\D)(2+5\D)(1-8\D)}{6(7+4\D)(5+8\D)}P_{1,3}, 
\ee
where $P_{1,3}$ is defined in \eqref{P13}. 
\\

In Table \ref{tab-mini-OPEs}, we compare the exact OPE coefficients of some 2d minimal models. 
We can see the OPE coefficients of the stress tensor $T$ are smaller than those of the leading scalar $\phi_2$, 
which explains the good estimates for the 2d CFTs in the toy example in section \ref{toy-ex}. 
There is a hierarchy in the OPE coefficients of the leading and subleading spin-2 operator $\{T,\,\phi_3\}$. 
From the ratios of the OPE coefficients, 
we expect the 2d Lee-Yang CFT can be easily truncated 
and the estimates for the 2d unitary minimal models are more accurate when $m$ is small. 
\\

\begin{table}[h!]
\centering
 \begin{tabular}{||c|| c c c c c c||} 
 \hline
 			& LY 	& m=3	&  m=4	& m=5  	&  m=6		& m=7  		\\ [0.5ex] 
 \hline
 \hline
$P_2$		& -3.7 	& 0.25	& 0.37	& 0.45	&	0.50		& 	0.53		\\
 \hline
$P_T/P_2$	& 0.010 	& 0.13   	& 0.15	& 0.17	&	0.19		& 	0.21	 	\\
 \hline
$P_3/P_T$	& 0		& 0   		& 0.046	& 0.070	&	0.082	& 	0.091	\\
 \hline
\end{tabular}
\caption{
The exact values or ratios of the OPE coefficients $P_2,\, P_T,\, P_3$ in several 2d minimal models \cite{DiFrancesco:1997nk}. 
LY stands for the 2d Lee-Yang CFT. 
The parameter $m$ indicates 2d unitary minimal models $\mathcal M(m+1,m)$. 
We only show two significant figures. 
$|P_T|$ are always smaller than $|P_2|$, 
while $|P_3|$ are much smaller than $|P_T|$. 
$P_3$ vanishes in the 2d Lee-Yang and Ising CFTs. 
 }
\label{tab-mini-OPEs}
\end{table}

In section \ref{2d-min-fr}, we consider the minimal fusion rule in the 2d CFTs
\be
\phi_1\times \phi_1=I+\phi_2+T,
\ee
where $\phi_2$ is assumed to be a scalar operator and $T$ is the stress tensor. 
Instead of scanning the complete two dimensional parameter space $\{\D_1,\,\D_2\}$, 
we simply set the spectral data of $\phi_1,\,\phi_2$ to (\ref{spectral-2d-m-1}, \ref{spectral-2d-m-2}), 
so the parameter space becomes one dimensional 
\footnote{We leave the complete scanning of 
the two-dimensional parameter space $\{\D_1,\,\D_2\}$ for future study.}.
Then we compare the estimates of the OPE coefficients $P_2,\, P_T$ 
from the truncation ansatz with the exact results \eqref{OPE-2d-m}.

The OPE coefficient  $P_3$ of the spin-2 operator $\phi_3$ has some interesting features. 
It vanishes at $\D=-2/5,\,-1,\,0,\,1/8$. 
These zeros are related to 2d minimal models $\mathcal M(5,\,2)$ 
and $\mathcal M(m+1,\,m)$ 
with $m=0,\,2,\,3$. 
\footnote{
When $\D=0$ or $\D=-1$, the OPE coefficient of the stress tensor $P_T$ also vanishes, 
so these two cases are in some sense trivial or unphysical. 
In the first case  ($m=2$), 
$\phi_{1,2}$ becomes an identity operator ($\D=0$) and the 4-point function reduces to 1  
\be
G(u,v)\big |_{\D\rightarrow 0}=1.
\ee 
In the second case ($m=0$), the scaling dimensions are $\D_1=-1$ and $\D_2=-2$.  
The 4-point function of $\phi_1$ reduces to
\be
G(u,v)|_{\D\rightarrow -1}=\frac 2 3 [1+u^{-1}+(u/v)^{-1}],
\ee
so we have a non-unitary free theory $\mathcal L\sim \phi_1\,\Box^2\phi_1$ 
and $\phi_2$ is a composite operator $\phi_2\sim \phi_1^2$. }
In particular, $\mathcal M(5,\,2)$ and $\mathcal M(4,\,3)$ correspond to 
the 2d Lee-Yang CFT and the 2d Ising CFT. 
In other words, the two physical CFTs can be identified from the decoupling of the subleading operator $\phi_3$. 
In section \ref{oper-decp}, we investigate this phenomenon in our truncation framework 
by adding operators to the minimal fusion rule. 

\subsection{Minimal fusion rule: $\phi_1\times \phi_1=I+\phi_2+T$}\label{2d-min-fr}
Let us consider the minimal fusion rule
\be
\phi_1\times \phi_1=I+\phi_2+T
\ee
in the context of 2d CFTs. 
The identity operator and the stress tensor are in the same twist family $\{\t_0=0\}$. 
The scalar operator $\phi_2$ is in the second twist family $\{\t_2=\D_2\}$. 

The conformal invariant part of the 4-point function becomes
\be
G(u,v)=v^{-\D_1}\sum_{i,j=0,2}\,\sum_{m,n=0}^M c_{i,j}^{m,n} u^{\t_i/2+m} v^{\t_j/2+n},
\ee 
where $M$ is the twist cutoff. 

We consider different approximation schemes:
\begin{itemize}
\item
At the zeroth order approximation, we set the twist cutoff as 
\be
M^{(0)}=0.
\ee 
Since we have one more unknown $P_T$ than the toy example in section \ref{toy-ex}, 
we need one more equation. 
We increase the descendant cutoff of the twist family $\{\t_0=0\}$ to $K_0=2$, 
because the new equation contains $P_T$. 
The solution of the linear system is
\be
P_2^{(0)}=\frac {4\D_1}{3\D_2-4\D_1},\quad
P_T^{(0)}=\frac 1 4 \D_1(\D_2-2\D_1).
\label{OPE-minimal-2d-0}
\ee
The expression of $P_2$ is the same as the result \eqref{P-ex} in the toy example. 
Let us substitute $\D_1,\,\D_2$ with \eqref{spectral-2d-m-1}, 
then we have
\be
P_2^{(0)}=\frac {2\D}{1+2\D},\quad
P_T^{(0)}=\frac 1 6 \D(1+\D).
\label{2d-P2-PT-0}
\ee

\item
Then we consider the first order approximation where the twist cutoff becomes 
\be
M^{(1)}=1. 
\ee
An appropriate descendant cutoff is
\be
K^{(1)}=3.
\ee
The number of unknowns is 12 and the number of equations is $(4+2)\times 2=12$, 
so we can solve the linear system. 
Using the minimal model values \eqref{spectral-2d-m-1}, the solutions become
\be
P_2^{(1)}=
-\frac{12\D (-2 + \D)  (5 + 8 \D) (-13 + 7 \D + 2 \D^2)}
{-670 - 1435 \D +  492 \D^2 + 884 \D^3 -   424 \D^4 + 192 \D^5},
\label{2d-P2-1}
\ee
\be
P_T^{(1)}=
\frac{\D (1 + 4 \D) (-54 - 137 \D + 63 \D^2 + 69 \D^3 + 4 \D^4)}
{-670 - 1435 \D +  492 \D^2 + 884 \D^3 -   424 \D^4 + 192 \D^5}.
\label{2d-PT-1}
\ee
The denominators coincide because they are from the same determinant. 
Surprisingly, $P_T^{(1)}$ gives the exact values of the central charges of the 2d Lee-Yang and Ising CFTs
\footnote{In \cite{ElShowk:2012hu}, as a warm-up example of the extremal functional method, 
similar estimates of the two OPE coefficients were discussed 
by numerically solving the truncated crossing equations. 
If we do not assume \eqref{spectral-2d-m-1} and 
use the same input $\D_1=0.125,\,\D_2=1.03$ from \cite{ElShowk:2012hu}, 
the first order approximation of $P_2$ is the same, 
but the approximate central charge $c\sim 0.48$ 
is closer to the exact value $c=0.5$ 
than the estimate in \cite{ElShowk:2012hu} which is $c\sim 0.45$.}.

\item
If we go to the second order 
\be
M^{(2)}=2,
\ee 
the appropriate descendant cutoff is 
\be
K^{(2)}=5.
\ee 
We need to neglect one equation so that the numbers of equations and unknowns match. 
We consider two schemes: 
in the first one, we neglect the last equation ($m=M,\, k=K$) in the twist family $\{\t_0\}$; 
in the second one, we omit the last equation ($m=M,\, k=K$) in the twist family $\{\t_2\}$. 
The solutions turn out to be the same rational functions of $\D$. 
They are denoted by
\be
P_2^{(2)},\quad P_T^{(2)}.
\ee
Their explicit expressions are much more involved, so we will not write them down. 
The estimates for $P_2$ are improved, 
but the estimates for $P_T$ are less accurate. 
The scaling dimensions associated with the descendant cutoff 
are already much higher than those of the operators in the truncated fusion rules, 
so we will not go to higher orders. 
\end{itemize}

In Table \ref{tab-mini-P2} and Table \ref{tab-mini-PT}, we compare the estimates at different approximation orders 
with the exact results. 
Using the minimal fusion rule, 
the first order approximations (\ref{2d-P2-1}, \ref{2d-PT-1}) already give 
estimates of the OPE coefficients with less than 10\% relative errors. 
\\

\begin{table}[h!]
\centering
 \begin{tabular}{||c|| c c c c c c|| c||} 
 \hline
 $P_2$	& LY 	& m=3	&  m=4	& m=5  	&  m=6	& m=7  	& $\eta$\\ [0.5ex] 
 \hline
 \hline
0th		& -4 		& 0.2	 	& 0.286	& 0.333	&	0.364	& 	0.385	& $< 30$\%	 \\
 \hline
1st		& -3.70 	& 0.243   & 0.353	& 0.415	&	0.454	& 	0.481	 & $< 10$\%\\
 \hline
2nd		& -3.62	& 0.246   & 0.364	& 0.434	&	0.479	& 	0.511	 & $< 5$\%\\
 \hline
exact	& -3.65	& 0.25	& 0.373	& 0.446	&      0.496	&	0.532	& \\ 
 \hline
\end{tabular}
\caption{The estimates of the OPE coefficient $P_2$ at different approximation orders and their exact values. 
LY stands for the 2d Lee-Yang CFT. 
The parameters $m$ indicate 2d unitary minimal models $\mathcal M(m+1,m)$. 
We only show three significant figures. 
The estimates are more accurate at high orders. 
The relative errors $\eta_2^{(i)}$ are defined as $|1-P_2^{(i)}/P_2^{\text{\,exact}}|$. 
The relative errors increase as we consider larger $\D$.
 }
\label{tab-mini-P2}
\end{table}

\begin{table}[h!]
\centering
 \begin{tabular}{||c|| c c c c c c ||c||} 
 \hline
$P_T$ 	& LY 	& m=3	&  m=4	& m=5  	&  m=6	& 	m=7 		& $\eta$\\ [0.5ex] 
 \hline
 \hline
0th		& -0.04 	& 0.023	& 0.04	& 0.052	&	0.061	& 	0.068 	&  $< 50$\% 
(not LY)	 \\
 \hline
1st		&-0.018 	& 0.016   	& 0.030	& 0.042	&	0.052	& 	0.059	&	$< 10$\%	 \\
 \hline
2nd		& -0.021	& 0.017   	& 0.034	& 0.048	&	0.061	&	0.072	&	$< 35$\% \\
 \hline
exact	& -0.018	& 0.016	& 0.029	& 0.039	&      0.048	&	0.055	&\\ 
 \hline
\end{tabular}
\caption{The estimates of the OPE coefficient $P_T$ at different approximation orders and their exact values. 
LY stands for the 2d Lee-Yang CFT. 
2d unitary minimal models $\mathcal M(m+1,m)$ are denoted by $m$. 
We only show two significant figures. 
The optimal estimates are from the first order approximations. 
Surprisingly, the approximate equation $P_T^{(1)}$ in \eqref{2d-PT-1} generates the exact values 
in the cases of the 2d Lee-Yang CFT and the 2d Ising CFT. 
We suspect this is related to the decoupling of the spin-2 operator $\phi_3$. 
The relative errors $\eta_T^{(i)}$ are defined as $|1-P_T^{(i)}/P_T^{\text{\,exact}}|$. 
In the first order approximation, 
the relative errors increase for larger $\D$. 
However, the second order approximation is less accurate, 
which indicates the importance of subleading operators. 
 }
\label{tab-mini-PT}
\end{table}

\newpage
The truncated fusion rule is not restricted to the operator products of the lowest scalar operators. 
In fact, in many 2d minimal models, $\phi_{1,2}$ is not the primary operator with the lowest scaling dimension.  

\subsection{Operator decoupling}\label{oper-decp}
In the interpolating solution between 2d minimal models, 
the OPE coefficient $P_3$ of the subleading spin-2 operator ($\t_3=\D_2$) vanishes 
when $\D$ takes some special values. 
The zeros at $\D=-2/5$ and $\D=1/8$ correspond to the 2d Lee-Yang CFT and the 2d Ising CFT. 

We want to study the operator decoupling phenomenon in the truncation framework. 
It is necessary to introduce the subleading operators to the truncated fusion rules. 
The decoupling of the subleading spin-2 operator $\phi_3$ provides alternative definitions of the 2d Lee-Yang CFT and the 2d Ising CFT. 

\subsubsection{2d Lee-Yang CFT}
Let us introduce the subleading spin-2 operator $\phi_3$ to the truncated fusion rule
\be
\phi_1\times\phi_1=I+\phi_2+T+\phi_3.
\ee

\begin{figure}[h!]
\begin{center}
\includegraphics[width=12cm]
{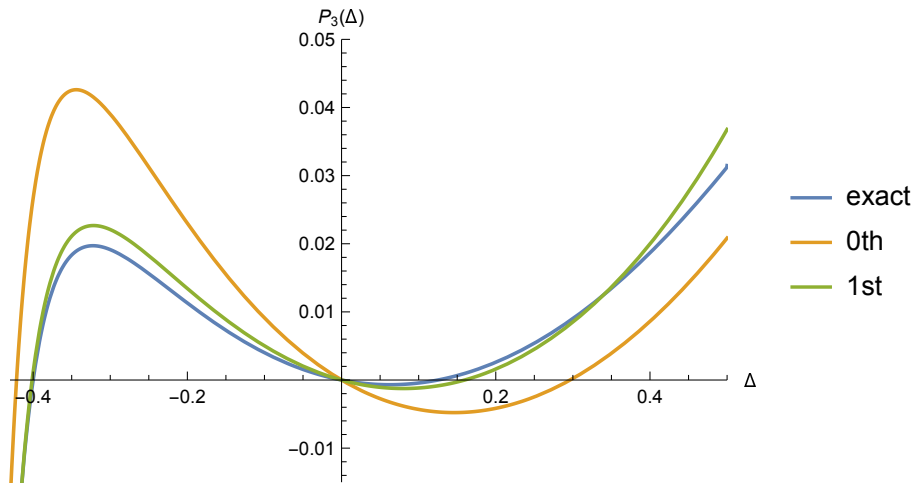}
\caption{The exact values (blue) of the OPE coefficient $P_3(\D)$ as a function of the scaling dimension $\D$ 
of $\phi_{1,2}$ 
and its estimates in the zeroth (orange) and the first (green) order approximations. 
The qualitative behavior of the first order estimate is close to the exact function. 
The Lee-Yang root of the first order estimate is rather accurate. 
} 
\label{fig-operdec-1}
\end{center}
\end{figure}

At the zeroth order, we set the twist cutoff as $M^{(0)}=0$ 
and the corresponding descendant cutoff is $K^{(0)}=2$. 
At the first order, the twist cutoff is $M^{(1)}=1$ 
and the appropriate descendant cutoff is $K^{(1)}=3$. 
In both cases, we need one more equation beyond the descendant cutoff. 
We always use the first equation beyond the descendant cutoff $\{m=0,\,k=K+1\}$ 
in the second twist family $\{\t=\D_2\}$. 
We choose the second twist family because these equations involve $P_3$, 
then the solutions of $P_3$ are more accurate. 
In Figure \ref{fig-operdec-1}, we compare the exact function of $P_3(\D)$ with the first-order and the second-order estimate functions. 

According to the first order solution, the Lee-Yang value is determined by the polynomial equation
\be
130 - 359 \D - 2511 \D^2 - 
 2099 \D^3 - 128 \D^4 + 
 192 \D^5=0,
 \ee
and the approximate Lee-Yang root is
\be
\D_{\text{2d Lee-Yang}}^{\text{estimate}}\sim -0.401,
\ee
where we only show three significant figures. 
The first order estimate of the Lee-Yang root is close to the exact value
\be
\D_{\text{2d Lee-Yang}}^{\text{exact}}=-0.4.
\ee

In Figure \ref{fig-operdec-1}, 
The first order estimate $P_3^{(1)}$ matches with the qualitative behavior of the exact solution. 
The estimate for the Ising value is $\D\sim 0.16$, which is not far from the exact value $\D=0.125$. 
To improve the Ising estimate, we need to introduce the higher operators.

\subsubsection{2d Ising CFT}
To obtain a more accurate Ising root, 
we introduce one more primary operator $\phi_4$ to the truncated fusion rule
\be
\phi_1\times\phi_1=I+\phi_2+T+\phi_3+\phi_4.
\ee

The possible choices for $\phi_4$ are
\be
\{\t_4,\,l_4\}=\{4,\,0\},\,\{0,\,4\},\,\{\D_2+4,\,0\},\,\{\D_2,\,4\}.
\ee
From the exact solution, we know the third operator $\{\t=\D_2+4,\,l=0\}$ also decouples at the 2d Ising point ($\D=1/8$), 
so we will not consider this possibility. 
The OPE coefficients of the other operators are in the same order of magnitude. 
\footnote{In principle, we should add all of them, but in this preliminary study we consider the minimal modification. }

To improve the accuracy of the 2d Ising root, 
we increase the twist cutoff to the second order ($M=2$). 
The number of unknowns is 25. 
The appropriate descendant cutoff is $K=5$ 
and the number of equations is $(6+4+2)\times 2=24$, 
so we need one more equation. 
We again use the first equation beyond the descendant cutoff ($m=0, k=K+1$) 
in the second twist family $\{\t=\D_2\}$. 
The solutions of $P_3$ are the same for the choices $\{\t_4=4,\,l_4=0\}$ and $\{\t_4=\D_2,\,l_4=4\}$. 
As a result, we have two independent solutions corresponding to 
\be
\{\t_4,\,l_4\}=\{4,\,0\},\, \{0,\,4\}.
\ee  

\begin{figure}[h!]
\begin{center}
\includegraphics[width=12cm]
{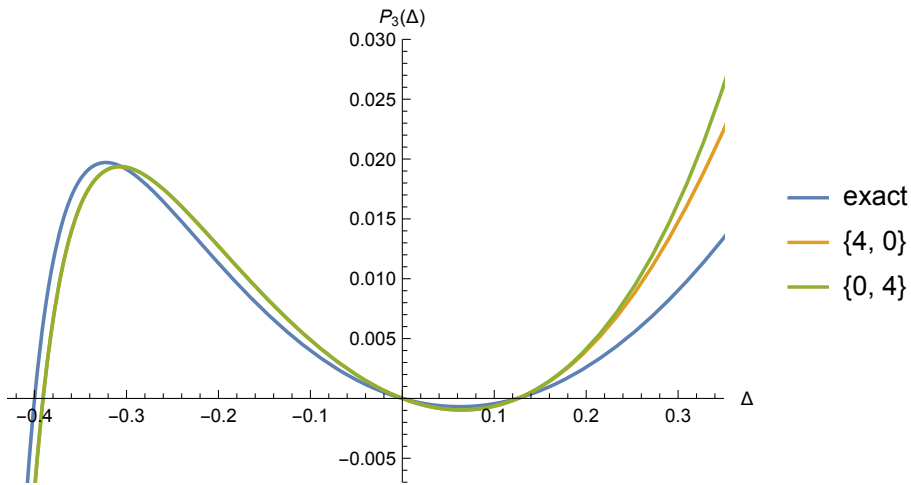}
\caption{The exact values (blue) of the OPE coefficient $P_3(\D)$ as a function of the scaling dimension $\D$ 
and its second order estimates with different $\phi_4$ operators: $\{\t_4,\,l_4\}=\{4,\,0\}$ (orange), $\{\t_4,\,l_4\}=\{0,\,4\}$ (green). 
The two approximate functions almost coincide when $\D$ is smaller than the Ising root. 
The estimate functions are close to the exact function between $\D=-0.05$ and $\D=0.15$. 
The Ising roots are rather accurate in the second order approximations, 
but the Lee-Yang roots are less accurate than the first order result. 
} 
\label{fig-operdec-2}
\end{center}
\end{figure}

In Figure \ref{fig-operdec-2}, the exact function and 
two second order approximate functions of $P_3(\D)$ are presented. 
The Ising roots of the two estimate function are given by the same polynomial equation
\be
39504 \D + 280496 \D^2 + 
 331542 \D^3 + 91875 \D^4 - 
 45629 \D^5 - 25992 \D^6 - 
 1728 \D^7 + 512 \D^8=10080.
\ee
The numerical value of the approximate Ising root is
\be
\D_{\text{2d Ising}}^{\text{estimate}}\sim 0.1257,
\ee
where we only show 4 significant figures. 
The estimate value is close to the exact value
\be
\D_{\text{2d Ising}}^{\text{exact}}=0.125.
\ee
From the decoupling of the subleading operator $\phi_3$, 
we are able to obtain a rather accurate estimate of the 2d Ising scaling dimension 
in the truncation framework.

\section{CFTs in various dimensions}\label{GenD-CFT}
In this section, we investigate CFTs in general dimensions in the truncation framework. 
In section \ref{mini-general-d}, we derive the approximate relations corresponding to the minimal fusion rule. 
Then we examine these relations in the canonical free CFTs and the Wilson-Fisher CFTs. 
We also apply these relations to the Lee-Yang CFTs and the Ising CFTs in various dimensions. 

\subsection{Minimal fusion rule: $\phi_1\times \phi_1=I+\phi_2+T$}\label{mini-general-d}
We mainly consider the minimal fusion rule
\be
\phi_1\times \phi_1=I+\phi_2+T,
\ee
where $I$ is the identity operator, $\{\phi_1,\,\phi_2\}$ are primary scalar operators and $T$ is the stress tensor. 
Let us assume that only the identity operator has a vanishing twist
\be
\t_I=0.
\ee
Then, from crossing symmetry, 
the identity term leads to a double-twist family $\{\t=2\D_1\}$ in the conformal partial wave expansion of $G(u,v)$
\be
G(u,v)=1+\left(\frac u v\right)^{\D_1}+u^{\D_1}+\dots
\ee

In the crossing symmetric function, we have two universal crossing-symmetric building blocks
\be
H(u,v)=(v^{\D_1}+u^{\D_1})+u^{\D_1}v^{\D_1}+\dots
\ee
The 4-point function of the fundamental scalar in a free theory 
contains only the universal part. 
The general form of the crossing symmetric function becomes
\be
H(u,v)=v^{\D_1}+u^{\D_1}+\sum_{i,j}\sum_{m,n=0}^M c_{i,j}^{m,n}\,
u^{\t_i/2+m} v^{\t_j/2+n},
\ee
\be
i,j=0,\,2, \,T,\quad
(i,m)\leq (j,n),
\ee
where the double twist family corresponds to $i=0$, namely
\be
\t_0=2\D_1. 
\ee 

The coefficient of the second universal crossing symmetric building block indicates 
$c_{0,0}^{0,0}=1$.
However, since we only use the crossing equation \eqref{crossing-1}, 
we will promote $c_{0,0}^{0,0}$ to a free parameter.

Due to the non-degeneracy of the twist-0 operator in the twist spectrum, 
there is only one twist-0 term $v^{\D_1}$, corresponding to the identity operator. 
To simplify the notation, we use $0$ to indicate the double-twist family 
and hopefully this notation will not lead to confusion. 

Let us derive the estimate equations of the OPE coefficients $P_2,\, P_T$ 
using the minimal fusion rule: 
\begin{itemize}
\item
At the zeroth order, we set the twist cutoff as 
\be
M^{(0)}=0. 
\ee
The appropriate descendant cutoff is
\be
K^{(0)}=2. 
\ee
The number of equations is slightly larger than the number of unknowns. 
We need to omit one equation. 
\begin{enumerate}
\item
In the first scheme, we omit the last equation ($k=K$) in the twist family $\{\t=\D_2\}$. 
The solutions of the OPE coefficients are
\be
P_2^{(0,1)}=\frac {4\D_1}{3\D_2-4\D_1},\quad
P_T^{(0,1)}=\frac {\D_1\D_2(\D_2-2\D_1)}{4(\D_2+2-d)}.
\label{OPE-genD-0-1}
\ee
When $d=2$, the OPE coefficients reduce to the 2d results \eqref{OPE-minimal-2d-0}.
\item
In the second scheme, we neglect the last equation ($k=K$) in the double twist family $\{\t=2\D_1\}$. 
The zeroth order estimates of the OPE coefficients are
\be
P_2^{(0,2)}=\frac {4\D_1}{3\D_2-4\D_1},\quad
P_T^{(0,2)}=\frac {\D_1\D_2(\D_2-2\D_1)
(4\D_1-2\D_2+4\D_1\D_2-3\D_2^2)}
{8(2\D_1+2-d)(4\D_1-3\D_2)(\D_2+1)}.
\label{OPE-genD-0-2}
\ee
The $d=2$ limit of $P_T$ is different from that in \eqref{OPE-minimal-2d-0}.
\item
In the third scheme, we omit the last equation in the twist family $\{\t=d-2\}$. 
However, the linear system has no solution 
because the equations do not contain $P_T$. 
This scheme is inconsistent. 
\end{enumerate}
We have two different schemes for the zeroth order approximation. 
The estimate equation $P_T^{(0,1)}$ in \eqref{OPE-genD-0-1} works better than 
the more complicate equation $P_T^{(0,2)}$ in \eqref{OPE-genD-0-2}. 
In the examples, we only present the zeroth order estimates of \eqref{OPE-genD-0-1} 
and then focus on the first order estimates. 

\item
At the first order, we set the twist and the descendant cutoffs as 
\be
M^{(1)}=1,\quad K^{(1)}=4. 
\ee
The number of unknowns is 23, while the number of equations is 
$(5+3)\times 3=24$. 
To obtain a solution, we need to omit at least one equation. 
However, no general solution is found if only one equation is neglected, 
which means not all the equations are compatible with each other. 
This is related to the fact that we can solve $P_2,\, P_T$ using less than $23$ equations. 
We consider three different choices:
\begin{enumerate}
\item
In the first case, we decrease the descendant cutoff of the double twist family $\{\t_0=2\D_1\}$ to $K^{(1)}_0=3$. 
The solutions are denoted by 
\be
P_2^{(1,1)},\quad P_T^{(1,1)}.
\label{OPE-genD-1-1}
\ee

\item
In the second case, we choose a lower descendant cutoff for the twist family $\{\t_2=\D_2\}$: $K^{(1)}_2=3$. 
The solutions are given by 
\be
P_2^{(1,2)},\quad P_T^{(1,2)}.
\label{OPE-genD-1-2}
\ee

\item
In the third case, we lower the descendant cutoff of the twist family $\{\t_T=d-2\}$ to $K^{(1)}_T=3$. 
The solutions are  
\be
P_2^{(1,3)},\quad P_T^{(1,3)}.
\label{OPE-genD-1-3}
\ee

\end{enumerate}
In each scheme, we use only 22 equations to solve the linear system. 
Although one $c_{i,j}^{m,n}$ remains arbitrary, 
the solutions of the approximate OPE coefficients $P_2,\,P_T$ are fixed. 
The free parameter can be determined using an appropriate equation beyond the descendant cutoff. 

Due to the large number of unknowns in the linear system, 
the first order solutions are fractions of two high order polynomials. 
We will not write down their explicit expressions. 
\end{itemize}

We mainly examine the first order estimates in the concrete examples
\be
P_2^{(1,i)},\quad P_T^{(1,i)},\quad i=1,2,3.
\ee 
Although we assume the twist-0 operators are non-degenerate, 
we also test these approximate relations in 2d CFTs. 
They give accurate estimates for the 2d OPE coefficients as well, 
which confirms our previous statement that the CFT data are insensitive to the choice of our basis functions. 

\subsection{Canonical free scalar theory}
Let us consider the canonical free scalar field theory in general dimensions. The Lagrangian reads
\be
\mathcal L=\frac 1 2 (\pa \phi_1)^2.
\ee
In the 4-point function of $\phi_1$, the conformal invariant part $G(u,v)$ is
\be
G(u,v)=1+\left(\frac u v\right)^{\frac {d-2}2}+u^{\frac {d-2}2}
\ee
and the associated crossing symmetric function is
\be
H(u,v)=v^{\frac {d-2}2}+u^{\frac {d-2}2}+(u\,v)^{\frac {d-2}2}. 
\ee
Note that we do not assume $d$ is an integer. 

We can decompose $G(u,v)$ into conformal blocks
\be
G(u,v)=1+P_2\, F_{\t_2,\,l_2}(u,v)+P_T\, F_{\t_T,\,l_T}(u,v)+\dots,
\ee
where the fusion rule is
\be
\phi_1\times \phi_1=I+\phi_2+T+\dots,
\ee
and $\phi_2$ is a composite operator
\be
\phi_2\sim (\phi_1)^2.
\ee

The spectral data of the low-lying operators are
\be
\{\D_1=(d-2)/2,\,l_1=0\},\quad
\{\D_2=d-2,\,l_2=0\},\quad
\{\t_T=d-2,\,l_T=2\}. 
\ee
All exchanged operators are in the double-twist family $\{\t=2\D_1\}$. 

The exact values of $P_2,\, P_T$ as functions of the spacetime dimension $d$ are
\be
P_2=2,\quad P_T=\frac {d( d-2)^2}{16 ( d-1)}. 
\ee

Let us check the estimates of the OPE coefficients by substituting 
\be
\D_1=\frac 1 2 (d-2),\quad\D_2=d-2
\ee 
into the approximate OPE coefficients from the truncation procedure. 

The zeroth order estimates \eqref{OPE-genD-0-1} give
\be
P_2^{(0,1)}=2,\quad
P_T^{(0,1)}=\frac {(d-2)^2}{8}.
\ee
The estimate of the OPE coefficient $P_2$ are exact, but that of $P_T$ is not accurate. 
Now we move to the first order approximations. 
Surprisingly, all the first order estimates (\ref{OPE-genD-1-1}, \ref{OPE-genD-1-2}, \ref{OPE-genD-1-3}) give the exact values
\be
P^{(1,i)}_2=2,\quad P^{(1,i)}_T=\frac {d( d-2)^2}{16 ( d-1)}.
\ee
Note that we do not make the assumption that the spacetime dimension $d$ is an integer, so $d$ is a continuous parameter.
We do not consider generalized free theories because the stress tensor usually decouples 
and their fusion rules are not consistent with the minimal fusion rule.

\subsection{$\phi^{2n}$ Wilson-Fisher CFTs}
Since the first order approximations give the exact OPE coefficients of the free scalar theory, 
it is interesting to consider small deformations of the free CFTs. 
We now examine the Wilson-Fisher CFTs where the $\epsilon$-parameter 
can be considered as a small deformation parameter. 
Note that we do not introduce global internal symmetry, 
so there is only one fundamental scalar operator in each case.

For $\phi^4$ theory in $d=4-\epsilon$ dimensions \cite{Wilson:1971dc}, the conformal data of the low-lying operators are
\footnote{There is one more relevant operator, i.e. $\phi^4$, but its squared OPE coefficient is of higher order in $\epsilon$. We will not consider its contribution. } 
\be
\D_1\big|_{\phi^4}=1-\frac \epsilon 2+\mathcal O(\epsilon^2),\quad
\D_2\big|_{\phi^4}=2-\frac 2 3 \epsilon+\mathcal O(\epsilon^2)
\ee
\be
P^{\text{\,exact}}_2\big|_{\phi^4}=2-\frac 2 3 \epsilon+\mathcal O(\epsilon^2),\quad
P^{\text{\,exact}}_T\big|_{\phi^4}=\frac 1 3 -\frac {11}{36}\epsilon+\mathcal O(\epsilon^2).
\ee

The first order estimates (\ref{OPE-genD-1-1}, \ref{OPE-genD-1-2}, \ref{OPE-genD-1-3}) of the OPE coefficients gives
\be
P^{(1,1)}_2\big|_{\phi^4}=2-\frac {59}{90} \epsilon+\mathcal O(\epsilon^2),\quad
P^{(1,1)}_T\big|_{\phi^4}=\frac 1 3 -\frac {29}{108}\epsilon+\mathcal O(\epsilon^2),
\ee
\be
P^{(1,2)}_2\big|_{\phi^4}=2-\frac  {85}{126}  \epsilon+\mathcal O(\epsilon^2),\quad
P^{(1,2)}_T\big|_{\phi^4}=\frac 1 3 -\frac {209}{726}\epsilon+\mathcal O(\epsilon^2),
\ee
\be
P^{(1,3)}_2\big|_{\phi^4}=2-\frac {59}{90} \epsilon+\mathcal O(\epsilon^2),\quad
P^{(1,3)}_T\big|_{\phi^4}=\frac 1 3 -\frac {29}{108}\epsilon+\mathcal O(\epsilon^2),
\ee
where the results of the first and the third schemes are different at the $\epsilon^2$ order. 
If we keep only three significant figures of the numerical values, the OPE coefficients become
\be
P^{\text{\,exact}}_2\big|_{\phi^4}=2-0.667\, \epsilon+\mathcal O(\epsilon^2),\quad
P^{\text{\,exact}}_T\big|_{\phi^4}=0.333 -0.306\,\epsilon+\mathcal O(\epsilon^2),
\ee
\be
P^{(1,1)}_2\big|_{\phi^4}=2-0.656\, \epsilon+\mathcal O(\epsilon^2),\quad
P^{(1,1)}_T\big|_{\phi^4}=0.333 -0.269\,\epsilon+\mathcal O(\epsilon^2),
\ee
\be
P^{(1,2)}_2\big|_{\phi^4}=2-0.675\,  \epsilon+\mathcal O(\epsilon^2),\quad
P^{(1,2)}_T\big|_{\phi^4}=0.333 -0.276\,\epsilon+\mathcal O(\epsilon^2),
\ee
\be
P^{(1,3)}_2\big|_{\phi^4}=2-0.656\, \epsilon+\mathcal O(\epsilon^2),\quad
P^{(1,3)}_T\big|_{\phi^4}=0.333 -0.269\,\epsilon+\mathcal O(\epsilon^2).
\ee
The estimates of $P_2$ are close to the exact value, while those of $P_T$ are slightly less accurate.
\\

For $\phi^6$ theory in $d=3+\epsilon$ dimensions, the scaling dimensions of the low-lying scalars are
\be
\D_1\big|_{\phi^6}=\frac 1 2+\frac 1 2\epsilon +\mathcal O(\epsilon^2),\quad
\D_2\big|_{\phi^6}=1+\epsilon+\mathcal O(\epsilon^2),
\ee
then, to the $\epsilon^1$-order, the approximate equations (\ref{OPE-genD-1-1}, \ref{OPE-genD-1-2}, \ref{OPE-genD-1-3}) 
in the three schemes give the same estimates
\be
P^{(1,i)}_2\big|_{\phi^6}=2+0\,\epsilon+\mathcal O(\epsilon^2),\quad
P^{(1,i)}_T\big|_{\phi^6}=\frac 3 {32} +\frac {11}{64}\epsilon+\mathcal O(\epsilon^2),
\ee
where the $\epsilon^0$-order terms are the free OPE coefficients in 3d. 
To the $\epsilon^1$-order, the conformal data look like a canonical free scalar theory in $d=3+\epsilon$ dimensions. 
It will be interesting to check whether the corrections at the $\epsilon^1$-order are good estimates 
by computing $P_2,\,P_T$ in a different method. 
\\

We do not consider the $\phi^3$ theory in $d=6-\epsilon$ dimensions 
due to the singular contribution in the conformal block of the exchange primary $\phi$, 
which is induced by the conformal multiplet recombination $\Box \phi \sim \phi^2$. 
However, we can consider lower spacetime dimensions.
We study the Lee-Yang CFTs in $2\leq d<6$ dimensions in section \ref{LY-d}. 

\subsection{Lee-Yang CFTs in $d=2,3,4,5$ dimensions}\label{LY-d}
The non-trivial fixed points of the $\phi^3$ theory correspond to the Lee-Yang CFTs \cite{Fisher:1978pf}. 
In the Lee-Yang CFTs, the two scalars $\phi_1,\phi_2$ coincide, so their scaling dimensions are equal to each other
\be
\D_1=\D_2. 
\ee

The edge exponent $\s$ is related to the scaling dimension of $\phi_1$ by
\be
\s=\frac{\D_1}{d-\D_1}.
\ee

In Table \ref{tab-LY-P2} and Table \ref{tab-LY-PT}, we use the numerical values of the edge exponent in various dimensions 
to compute the estimates of the OPE coefficients $P_2,\,P_T$. 
In 2d, we use the exact value of $\s$ and compare with the exact OPE coefficients \cite{DiFrancesco:1997nk}. 
In higher spacetime dimensions, the input values are from \cite{Gliozzi:2013ysa, Gliozzi:2014jsa}, 
then in Table \ref{tab-LY-P2} we compare the estimate results with those in \cite{Gliozzi:2013ysa, Gliozzi:2014jsa}. 
If the input values of the edge exponent are accurate, 
the estimates of $P_T$ are the predictions of our truncation ansatz. 

\begin{table}[h!]
\centering
 \begin{tabular}{||c|| c |c c | c c | c|| } 
 \hline
			&  	d=2		& d=3	& d=3	&   d=4 	& d=4 		& $d=5$	 	\\ [0.5ex] 
\hline\hline
$\sigma_{\text{input}}$
			&  	-1/6 		& 0.076	&0.085	& 0.259	&0.2685 	& 	0.4105		 \\
\hline
$P_2^{(1,1)}$	&  	-3.63 	& -3.90	&-3.88 	& -2.82	&-2.67 	& 	-0.858	 \\ 
 \hline
$P_2^{(1,2)}$	&  	-3.63		& -3.90	&-3.88 	& -2.80	&-2.65 	& 	-0.807		 \\ 
 \hline
$P_2^{(1,3)}$	&  	-3.63	  	& -3.90	&-3.88 	& -2.83	&-2.67 	& 	-0.859		 \\ 
 \hline
$P_2^\text{\,exact/numerical}$	&  	-3.65		& -3.91	&-3.88(1)	& -2.86	&-2.72(1)	& 	-0.95(2)		 \\ 
 \hline
\end{tabular}

\caption{ 
The first order estimates of the OPE coefficient $P_2$ of the Lee-Yang CFTs in various dimensions and 
the exact value \cite{DiFrancesco:1997nk} or the numerical values from conformal bootstrap \cite{Gliozzi:2013ysa, Gliozzi:2014jsa}. 
We only show three significant figures of the estimates. 
The estimates are most accurate in three dimensional spacetime. 
The estimates in 5d seem to be the least accurate. 
 }
\label{tab-LY-P2}
\end{table}

\begin{table}[h!]
\centering
 \begin{tabular}{||c|| c |c c | c c | c|| } 
 \hline
			&  	d=2		& d=3	& d=3 	& d=4	& d=4 		& $d=5$	 	\\ [0.5ex] 
\hline\hline
$\sigma_{\text{input}}$
			&  	-1/6 		& 0.076	& 0.085	& 0.259	& 0.2685 	& 	0.4105		 \\
\hline
$P_T^{(1,1)}$	&  	-0.016 	& 0.0017	& 0.0024 	& 0.070	& 0.082 	& 	0.31	 \\ 
 \hline
$P_T^{(1,2)}$	&  	-0.016	& 0.0017	& 0.0024 	& 0.066	& 0.079	& 	0.24		 \\ 
 \hline
$P_T^{(1,3)}$	&  	-0.016	& 0.0017	& 0.0024 	& 0.071	& 0.083 	& 	0.31		 \\ 
 \hline
$P_T^\text{exact/numerical}$	
			&  	-0.018	& 0.0023	&$-$	& 0.13 	&	$-$   &   $-$			 \\ 
 \hline
\end{tabular}

\caption{
The first order estimates of the OPE coefficient $P_T$ in various dimensions and 
the exact value \cite{DiFrancesco:1997nk} or the numerical bootstrap values \cite{Gliozzi:2013ysa}. 
The input data are from \cite{Gliozzi:2013ysa, Gliozzi:2014jsa}.  
Some of the numerical values are absent. 
We only show two significant figures of $P_T$. 
The 2d estimate is close to the exact value, 
but the 3d and 4d estimates are less consistent with the bootstrap results in \cite{Gliozzi:2013ysa}. 
 }
\label{tab-LY-PT}
\end{table}

In 3d and 4d, we consider two different sets of input values from \cite{Gliozzi:2013ysa, Gliozzi:2014jsa}. 
In 3d, the estimates of $P_2$ match with the numerical results particularly well.
It seems the OPE coefficient $P_2$ in 3d is mainly determined by the edge exponent. 
\footnote{Note that $|P_T/P_2|$ is very small in 3d, 
which explains the accuracy of the 3d estimates of $P_2$.}
In contrast, the estimates of the stress tensor OPE coefficients are less close to the results in \cite{Gliozzi:2013ysa}. 
 
As the dimension of spacetime increases, 
the estimates are less consistent with the previous numerical results, 
which signals the importance of subleading operators. 
For instance, $|P_T|$ is much smaller than $|P_2|$ in 3d, 
but they are in the same order of magnitude in 5d.

\subsection{Ising CFTs in $d=2,3$ dimensions}
In this subsection, we want to test our estimate equations (\ref{OPE-genD-1-1}, \ref{OPE-genD-1-2}, \ref{OPE-genD-1-3}) in the Ising CFTs. 
The 2d Ising CFT corresponds to the 2d minimal model with $m=3$, 
so its exact CFT data are known. 
The 3d Ising CFT is the most prominent example of the modern numerical bootstrap method, 
where the low-lying conformal data are determined to high precision \cite{Simmons-Duffin:2016wlq}. 
\begin{table}[h!]
\centering
 \begin{tabular}{||c|| c | c  || } 
 \hline
			&  	d=2	& 	d=3 		\\ [0.5ex] 
\hline\hline
$(\D_1,\,\D_2)_{\text{input}}$
			&  	(0.125,\,1) 		& 	(0.5181,\, 1.413)	 \\
\hline
$(P_2,\, P_T)^{(0,1)}$	
			&  	(0.2,\, 0.023)	& 	(0.96,\, 0.17) 	 	\\ 
			\hline
$(P_2,\, P_T)^{(1,1)}$	
			&  	(0.254,\, 0.018)	& 	(1.12,\, 0.12) 	 	\\ 
 \hline
$(P_2,\, P_T)^{(1,2)}$	
			&  	(0.252,\, 0.018)	& 	(1.11,\, 0.11) 		 \\ 
 \hline
$(P_2,\, P_T)^{(1,3)}$	
			&  	(0.254,\, 0.019)	  & 	(1.12,\, 0.12)		 \\ 
 \hline
$(P_2,\, P_T)^{\text{exact/numerical}}$
			&  	(0.25,\, 0.016)	& 	(1.11,\, 0.11)		 \\ 
 \hline
\end{tabular}

\caption{ 
The estimates of the Ising OPE coefficient $P_2,\,P_T$ in 2d and 3d,
the 2d exact values \cite{DiFrancesco:1997nk} and the 3d numerical bootstrap values \cite{Simmons-Duffin:2016wlq}. 
We only show three significant figures of $P_2$ and two significant figures of $P_T$. 
The zeroth order equations are \eqref{OPE-genD-0-1} and 
the first order equations are (\ref{OPE-genD-1-1}, \ref{OPE-genD-1-2}, \ref{OPE-genD-1-3}). 
The first order estimates for the OPE coefficients of the 3d Ising CFT are rather accurate.
 }
\label{tab-Ising}
\end{table}

In Table \ref{tab-Ising}, we present the estimates of the OPE coefficients $P_2,\,P_T$ 
by the zeroth order equations \eqref{OPE-genD-0-1} and 
the first order equations (\ref{OPE-genD-1-1}, \ref{OPE-genD-1-2}, \ref{OPE-genD-1-3}). 
The estimates for the 3d Ising CFT are particularly accurate. 
The 2d estimates are slightly less accurate.

\section{Discussion}\label{discuss}
In this work, we develop a novel approach to study conformal field theories using the conformal bootstrap. 
This approach is different from the standard method. 
Our starting point is the crossing symmetric functions, 
so the non-trivial crossing equation \eqref{crossing-1} is manifestly solved. 
Let us emphasize that, after introducing the cutoffs, 
the truncated crossing symmetric functions are studied in the Euclidean regime where the OPE convergence is rapid. 

In this new perspective, we have a natural truncation ansatz in the crossing solution space. 
We focus on the minimal truncated fusion rule 
\be
\phi_1\times \phi_1=I+\phi_2+T,
\ee
and derive some relations for the CFT data of the truncated spectrum. 
From these approximate relations, 
one can estimate the OPE coefficients using the scaling dimensions of two scalar operators. 
For instance, if one measures the scaling dimensions of the two scalars $\phi_1,\phi_2$, 
we can predict the magnitude of the three point function coefficient of $<\phi_1\,\phi_1\,\phi_2 >$. 
The prediction should be accurate 
if the coefficients of other three point functions $<\phi_1\,\phi_1\,\mathcal O >|_{\mathcal O\neq \phi_2}$ are comparably small. 

In section \ref{GenD-CFT}, 
we test the first order approximate equations (\ref{OPE-genD-1-1}, \ref{OPE-genD-1-2}, \ref{OPE-genD-1-3}) 
in several CFTs in various dimensions. 
The estimates of the OPE coefficient $P_2$ are particularly accurate and 
the estimates of $P_T$ are consistent with the well-established results. 
In 2d CFTs, due to the absence of twist gaps, the structure of the crossing-symmetric functions is slightly different, 
so we have different estimate equations (see section \ref{2d-min-fr}). 
But the equations (\ref{OPE-genD-1-1}, \ref{OPE-genD-1-2}, \ref{OPE-genD-1-3}) 
also give rather accurate estimates of the 2d OPE coefficients. 
Therefore, the equations (\ref{OPE-genD-1-1}, \ref{OPE-genD-1-2}, \ref{OPE-genD-1-3}) are universal!
\\

In section \ref{oper-decp}, the 2d Lee-Yang and the 2d Ising CFTs are identified in our truncation framework 
based on the phenomenon of operator decoupling.  
It will be very interesting to investigate the decoupling of subleading operators analytically in 3d CFTs, 
which was observed in the numerical bootstrap study of the 3d Ising CFT \cite{El-Showk:2014dwa}. 

Let us stress that the estimate equations (\ref{OPE-genD-1-1}, \ref{OPE-genD-1-2}, \ref{OPE-genD-1-3}) 
work exceptionally well in 3d, i.e. in the 3d Lee-Yang CFT and the 3d Ising CFT.
In the traditional analytic methods, 3d CFTs are the least accessible 
because they are far from the free theories 
in the $\epsilon$-expansions and, unlike the 2d minimal models, they do not have exact solutions. 
The encouraging results suggest that we should further develop 
this approach to study the 3d CFTs  systematically.

To improve the truncation ansatz, we need to understand why sometimes 
higher order approximations or longer truncated fusion rules 
do not give better estimates. 
We think this is related to the fact that the truncation ansatz is not 
a perturbation procedure where one expands the results in terms of some small parameters. 
By higher order, we instead mean the cutoffs are increased 
and the space of crossing symmetric functions is enlarged. 
In addition, all the operators in the truncated fusion rules are treated equally without using the information 
that the leading operators are more important. 
\footnote{However, the estimate equations know the OPE coefficients of the leading operators should be larger. 
By leading operators, we mean operators of low scaling dimensions and low spins. }
As a result, after we introduce more operators to the fusion rules, 
the estimates for the leading OPE coefficients sometimes become less accurate, 
which can be traced back to the instability of the subleading operators.
An interesting direction to be investigated is to develop a perturbation procedure 
where the OPE coefficients of the subleading operators are the small expansion parameters. 
A byproduct may be a better control of the error estimations. 
Note that we do not attempt to assign error bars in the present work. 

It is also important to understand how to incorporate the O(N) models in our truncation framework. 
For example, in $\vec\phi^4$ theory, the OPE of two fundamental scalars $\phi_i,\, \phi_j$ 
involves more than one relevant scalars: 
a singlet and a traceless tensor operators, 
corresponding to the composite operators $\vec\phi^2$ and $\phi_i \phi_j - \d_{ij}\,\vec \phi^2/N$. 
In the minimal fusion rule, there is only one relevant scalar, 
so it seems that we need to extend the truncated fusion rule. 
The issue of mixed correlators also deserves investigation. 

It would be interesting to consider the truncation framework in different coordinate systems or even different representation. 
For example, it was showed in \cite{Hogervorst:2013sma} that the polar coordinate has better convergent properties. 
In the Mellin representation, the conformal blocks, which are infinite series in the position space, 
are instead related to polynomials of finite degrees \cite{Mack:2009mi,Penedones:2010ue,Fitzpatrick:2011ia,Paulos:2011ie,Dolan:2011dv,Fitzpatrick:2012cg}. 

An ambitious direction is the extension to general quantum field theories without conformal symmetry. 
In the truncation procedure, we do not consider the high level descendants, 
but they play an important role in preserving conformal invariance. 
It seems the rapid OPE convergence is more crucial than conformal symmetry in this bootstrap method. 
It is natural to extend the inverse bootstrap approach to general QFTs. 
\\

For practical reasons, we mainly discuss the approximate results from the truncation procedure in this work. 
Now we switch to the discussion of exact results. 
If some CFTs share the same general solution of the crossing equations, 
they are connected by continuous deformations of the spectral data $\{\D_i,\,l_i\}$ and the spacetime dimension $d$, 
while the OPE coefficients are determined by crossing symmetry. 
The fact that many physical CFTs are consistent with the same approximate relations 
seems to indicate that they are connected to each other.  
One example is the Wilson-Fisher fixed points 
which smoothly connect free CFTs with interacting $\phi^n$ theories
\footnote{In the works \cite{Rychkov:2015naa,Basu:2015gpa,Ghosh:2015opa,Raju:2015fza,Nii:2016lpa,Hasegawa:2016piv,Gliozzi:2016ysv,Gliozzi:2017hni}, 
some Wilson-Fisher fixed points are obtained by 
requiring smooth deformations from free theories. 
}, 
with the spacetime dimension $d$ being the deformation parameter. 
\footnote{Are different $\phi^n$ fixed points connected by the continuous deformation of $n$? 
The marginal deformation of a free theory $\mathcal L=\phi \Box^k\phi$ corresponds to $n=2d/(d-2k)$. 
Note that $d$ is fixed in this footnote. 
The interpolation between 2d minimal models \cite{Liendo:2012hy} may furnish an example. 
} 
Another example is the interpolation between the 2d minimal models 
where the deformation parameter corresponds to the central charge. 
In both cases, the unitary CFTs are smoothly connected by some non-unitary CFTs 
\footnote{See \cite{Hogervorst:2015akt} for the non-unitary nature of 
CFTs in non-integer spacetime dimensions.}. 

Then we have two interesting questions. Are the general solutions of the crossing equations unique? 
If not, can we classify them? 
To address these two questions, 
we need to understand better the topology of the space of crossing symmetric functions, 
i.e. the solution space of the crossing equations. 
In a connected region, we expect the OPE coefficients share a universal formula as a function of $\{d,\D_i,l_i\}$,  
but one should distinguish the spectral data of the operator under consideration 
from the others.  
\\

A different, but more physical, classification will be based on the shortest, consistent truncation of a given fusion rule. 
Since in this work we only consider CFTs with single fundamental scalar operator in the Lagrangian descriptions, 
it is natural that the fusion rules with single relevant scalar are in the same ``universality class". 
The unexpected feature is that the differences in discrete symmetries seem to be less crucial. 
Perhaps continuous symmetries are more important, as we have additional conserved currents. 
One of the simplest examples is the O(N) model discussed above.

\section{Acknowledgements}
I would like to thank Euihun Joung and Jin-Beom Bae for stimulating discussions. 
This work was supported by the National Research Foundation of Korea through the grant NRF-2014R1A6A3A04056670.

\newpage
\appendix
\renewcommand{\theequation}{\thesection.\arabic{equation}}
\addcontentsline{toc}{section}{Appendix}
\section*{Appendix}

\section{Series expansion of conformal blocks}\label{App-series-cb}
In this Appendix, we summarize some relevant properties of 
the series expansion of conformal blocks \cite{Dolan:2011dv,Ferrara:1973vz,Ferrara:1974nf,Ferrara:1974ny,Dolan:2000ut,Dolan:2003hv}. 
The twist and the spin  
of the exchanged primary operator $\mathcal O$ are denoted by $\t,l$. 
The scaling dimension $\D$ is given by
\be
\D=\t+l.
\ee 
We will use $\t$ and $l$ as the independent spectral data of the exchanged primary field.  

The conformal invariant part of a four-point function can be decomposed into a sum of conformal blocks. 
The conformal block of the primary $\mathcal O$ and its descendants has a double power series representation
\be
F_{\t,\,l}(u,v)=u^{\t/2}\sum_{m,n=0}^\infty b_{m,n}(\t,l,d)\, u^m (1-v)^n.
\ee

The series coefficients at the lowest twist order ($m=0$) have a closed form expression
\be
b_{0,n}(\t,l,d)=\frac {(\t/2+l-\D_{12}/2)_{n-l}\,(\t/2+l+\D_{34}/2)_{n-l}}{(n-l)!\,(\t+2l)_{n-l}},
\ee
where $\D_{12},\,\D_{34}$ are defined by the scaling dimensions of the external scalar operators
\be
\D_{12}=\D_1-\D_2,\quad
\D_{34}=\D_3-\D_4,
\ee
$(x)_n$ is the Pochhammer symbol
\be
(x)_n=\G(x+n)/\G(x),
\ee
and $b_{0,n}(\t,l,d)$ vanishes if $n-l$ is a negative integer. 
From the quadratic Casimir equation of conformal blocks \cite{Dolan:2003hv}, 
when $m>0$, 
the series coefficients $b_{m,n}$ satisfy a recursion relation
\ba
&&[m(2m+2n-d)+n(n-1)-l(l-1)+\t(2m+n-l)]b_{m,n}
\nn&=&
(\t/2+m+n-1-\D_{12}/2)(\t/2+m+n-1+\D_{34}/2)
\left(b_{m,n-1}+b_{m-1,n}\right)
\nn&&
+2 (n + 2) (n + 1)b_{m-1,n+2}
-(n + 1) ( 2 m+ 3 n + \t-\D_{12}+\D_{34})b_{m - 1, n + 1}\quad
\ea
with
\be
b_{m,n}=0,\quad \text{if}\quad n<0\quad \text{or}\quad n<l-2m.
\ee

From this recursion relation, 
the first two coefficients at each twist order also have closed form expressions
\be
b_{m,\,l-2m}=\frac {(-1)^m(-l)_{2m}} {m!\, (l+d/2-m-1)_m},
\ee
and
\be
b_{m,\,l-2m+1}=\frac 1 2 b_{m,\,l-2m}
(l+\t/2-m-\D_{12}/2+\D_{34}/2-\D_{12}\,\D_{34}\, \tilde b_{m,\,l-2m+1}),
\ee
with
\be
\tilde b_{m,\,l-2m+1}=\frac {2m(l-1)+(l+1)(d-2)+\t(2m-l-1)}{2(l - 2 m+1) (d-\t-2) (2l + \t)}.
\ee
If the external scalars are identical, 
the second coefficient has a simple expression
\be
b_{m,l-2m+1}\Big|_{\D_{12}=\D_{34}=0}=\frac {(-1)^m(-l)_{2m}} {2 (m!) (l+d/2-m-1)_m}(l+\t/2-m)\,.
\ee

At high twist orders, the first few coefficients at each twist order vanish because $l-2m<0$. 
Then the first non-zero coefficients correspond to $b_{m,0}$. 

To some extent, the spin $l$ can be considered as a continuous parameter.

\end{document}